\documentclass[preprint,prc,showpacs,preprintnumbers]{revtex4-1}
\usepackage{graphicx,latexsym,amssymb,amsmath,color,multirow,mathrsfs,booktabs,ifpdf}
\usepackage{dcolumn}
\usepackage{bm}
\usepackage{longtable}

\def\oc{$^{16}$O+$^{12}$C\ }
\def\ac{$\alpha+^{12}$C\ }

\def\cc{$^{12}$C+$^{12}$C\ }
\def\oo{$^{16}$O+$^{16}$O\ }
\def\AA{nucleus-nucleus\ }
\def\Atrans{$^{12}$C ($^{16}$O,$^{12}$C)$^{16}$O\ }
\begin{document}
\title{Direct and indirect $\alpha$ transfer in the elastic \oc scattering}
\author{Nguyen Tri Toan Phuc$^{1,2}$}
\author{Nguyen Hoang Phuc$^1$} 
\author{Dao T. Khoa$^1$}
\affiliation{$^1$ Institute for Nuclear Science and Technology, VINATOM \\ 
179 Hoang Quoc Viet, Cau Giay, Hanoi, Vietnam. \\
$^2$ VNUHCM-University of Science, Ho Chi Minh City, Vietnam}
\begin{abstract}
The extensive elastic \oc scattering data measured at low energies show consistently 
an oscillating enhancement of the elastic cross section at backward angles that is 
difficult to describe within the conventional optical model. Given the significant 
$\alpha$ spectroscopic factors predicted for the dissociation $^{16}$O$\to\alpha+^{12}$C 
by the shell model (SM) and $\alpha$-cluster model calculations, the contribution of the 
$\alpha$ transfer channels to the elastic \oc scattering should not be negligible, 
and is expected to account for the enhanced oscillation of the elastic cross section 
at backward angles. To reveal the impact of the $\alpha$ transfer, a systematic 
coupled reaction channels (CRC) analysis of the elastic \oc scattering has been 
performed where the multistep couplings between the elastic and inelastic scattering 
channels, the direct and indirect $\alpha$ transfer channels were treated explicitly, 
using the real optical potentials and inelastic scattering form factors determined 
by the double-folding model. We show that a consistent CRC description of the 
elastic \oc data at different energies can be obtained over the whole angular region, 
using the $\alpha$ spectroscopic factors determined recently in the large scale SM 
calculation. The present CRC results are, therefore, of interest not only for the 
nuclear scattering studies but also provide an important spectroscopic 
information on the cluster dissociation of $^{16}$O.

\end{abstract}
\date{\today}
\maketitle

\section{Introduction}
While the elastic heavy-ion (HI) scattering is usually dominated by the strong 
absorption \cite{Sa79,Bra97}, some light HI systems are weak absorbing and refractive 
enough for the appearance of the nuclear rainbow pattern at medium and large 
angles, which allows the determination of the real \AA optical potential (OP) 
with a much less ambiguity (see, e.g., the topical review \cite{Kho07r} 
for more detail). As discussed in a recent folding model analysis \cite{Kho16} of the 
elastic \cc and \oc scattering, there is a range of the refractive energies 
($10\lesssim E \lesssim 40$ MeV/nucleon for the incident $^{12}$C and $^{16}$O ions), 
where the nuclear rainbow pattern can be clearly observed. Although the \cc and \oo 
systems are strongly refractive, the rainbow pattern cannot be observed at 
$\theta_\text{c.m.}>90^\circ$ because of the boson symmetry of the two identical 
nuclei that leads to a rapidly oscillating elastic cross section around the angle
$\theta_\text{c.m.}=90^\circ$. The \oc system does not have the boson symmetry 
and was considered as a good candidate for the study of the nuclear rainbow 
\cite{Bra91}. For that purpose, several experiments have been performed to measure 
the elastic \oc scattering with high-precision, covering a wide range of energies 
($E_\text{lab}\approx 20-1503$ MeV) and a broad angular region (up to 
$\theta_\text{c.m.}>130^\circ$ at low energies) 
\cite{Rou85,Bra86,Vil89,Oglo98,Oglo00,Nico00,Bra01}.
Very interesting are the elastic \oc data measured at the HI cyclotrons of the 
Kurchatov institute and university of Jyv\"{a}skyl\"{a} 
\cite{Oglo98,Oglo00}, which were shown to exhibit a pronounced 
nuclear rainbow pattern \cite{Kho16,Oglo00}, and the low-energy data measured 
at the Strasbourg Tandem Vivitron \cite{Nico00}. The extensive optical 
model (OM) and folding model studies of the elastic \oc scattering 
\cite{Kho16,Kho94,Kho97,Kho00a,Bra88,Mich01,Ohku14-1} have shown unambiguously 
the nuclear rainbow pattern in this system. However, at low energies 
($E_\text{lab}\lesssim 132$ MeV) the smooth rainbow pattern at backward angles 
is strongly deteriorated by a quick oscillation of the elastic \oc cross section 
(see, e.g., the upper panel of Fig.~11 in Ref.~\cite{Kho16}). In the conventional 
OM using the empirical Woods-Saxon (WS) potentials, one could obtain a good 
description of the low-energy elastic \oc data only if an extremely small 
diffuseness ($a_V\lesssim 0.1$ fm) of the absorptive WS potential is used 
\cite{Nico00}. Such an abrupt shape of the absorptive WS potential is drastically 
different from the global systematics of the complex OP for the 
\oc system \cite{Bra97}. 

\begin{figure}[bht]\vspace*{0cm}
\includegraphics[width=0.9\textwidth]{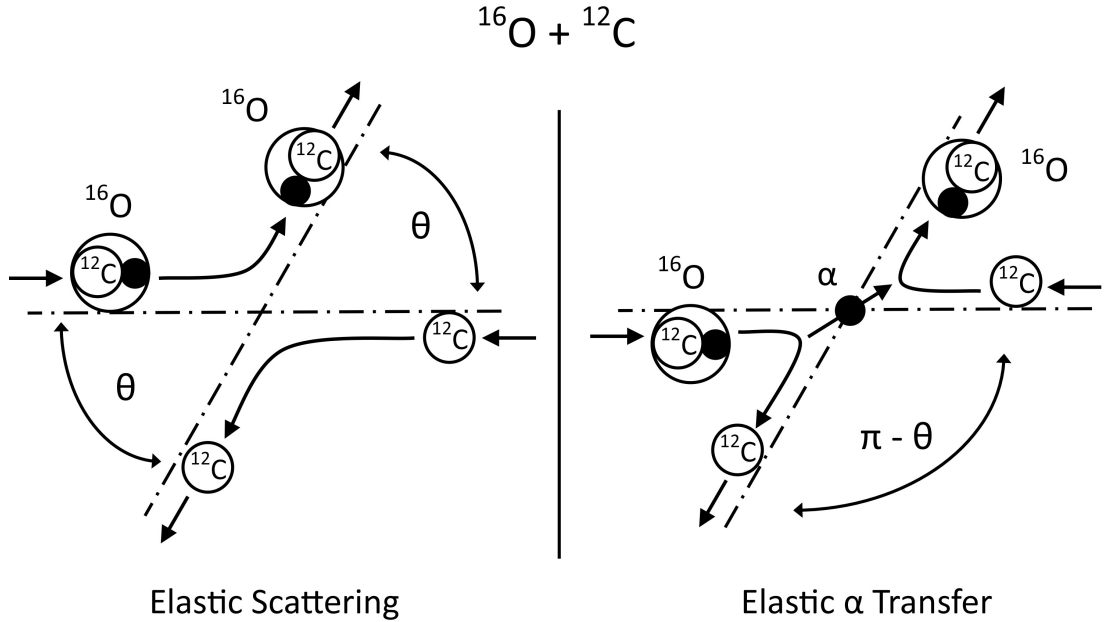}\vspace*{0cm}
 \caption{Kinematical illustration of the elastic scattering and elastic $\alpha$ 
 transfer processes in the \oc system.} \label{f1}
\end{figure}
In the present study, we focus on the high-precision elastic \oc scattering data, 
measured at low energies of 5 to 8 MeV/nucleon \cite{Oglo98,Oglo00,Nico00}, 
over the center-of-mass (c.m.) angles up to around $170^\circ$. Note that the OM 
analyses of the elastic \oc data available in the seventies and early eighties 
faced about the same problem \cite{Brau82,vOe75}, and a parity-dependent term 
was often added to the complex OP which was suggested by von Oertzen and Bohlen 
\cite{vOe75} as necessary to effectively account for the core exchange symmetry 
or the elastic $\alpha$ transfer between $^{16}$O and $^{12}$C (see Fig.~\ref{f1}). 
Guided by such a scenario, Szilner {\it et al.} \cite{Szi02} have analyzed the 
Strasbourg data measured at $E_{\rm lab}=100,\ 115.9$, and 124 MeV \cite{Nico00} within 
the coupled reaction channels (CRC) approach, explicitly taking into account the coupling 
between the elastic scattering and direct (elastic) $\alpha$ transfer channels. 
The observed oscillating cross sections at large angles were well described by these 
CRC results \cite{Szi02}, where the WS forms were used for the real OP and a weakly 
absorptive imaginary OP having ``standard" diffuseness of around $0.5-0.6$ fm. 
Similar analyses of the direct  $\alpha$ transfer in the elastic $^{16}$O+$^{12}$C 
scattering at energies near the Coulomb barrier were done in the distorted wave Born 
approximation (DWBA) \cite{Mor11,Hama11}. In these studies, parameters of the OP were 
fitted to the best OM description of the elastic cross section at forward angles, 
and the overall good description of the elastic \oc data was achieved with a DWBA 
amplitude of the direct $\alpha$ transfer added to the elastic scattering amplitude. 
The different elastic transfer processes in the elastic \oc scattering at low
energies were studied in the CRC calculation by Rudchik {\it et al.} 
\cite{Rud10}, and they were shown to contribute significantly to the elastic cross 
section. These CRC results show, however, a very weak contribution from the 
$\alpha$ transfer to the elastic cross section at backward angles. At variance 
with the conclusions made in the studies mentioned above, Ohkubo and Hirabayashi 
\cite{Ohku14-2} have proposed a completely different scenario for the quickly 
oscillating elastic \oc cross section at backward angles based on the results 
of their coupled channel (CC) analysis of the elastic \oc data measured at 
$E_{\rm lab}=115.9$ MeV. Namely, the oscillatory pattern at large angles was 
interpreted as the nuclear ``ripples" given by an interference between the elastic 
scattering wave and the external reflective wave (caused by the nuclear excitations 
taken into account in the CC calculation). Thus, the true physics origin 
of the oscillating enhancement of the elastic \oc cross section at backward angles, 
observed at energies $E_\text{lab}\lesssim 132$ MeV, is still under discussion.      
          
To explore the impact of the $\alpha$ transfer on the elastic \oc scattering, 
we have performed in the present work a systematic CRC analysis of the low-energy 
elastic \oc scattering data measured at the energies $E_{\rm lab}=100-124$ MeV 
by Strasbourg group \cite{Nico00}, the data measured at $E_{\rm lab}=132$ MeV 
by Kurchatov group \cite{Oglo98,Oglo00}, and the interesting data measured 
at $E_{\rm lab}=300$ MeV by Brandan {\it et al.} \cite{Bra01} including the data
points at the most backward angles that could not be described so far in the 
standard OM analysis. A proper choice of the OP for both the \oc and \cc systems 
is vital for the CRC analysis of the elastic \oc scattering at low energies. 
The real OP for these systems at the refractive energies has been shown to be 
well described by the (mean-field based) double-folded potential \cite{Kho16}. 
In a smooth continuation to lower energies, the recently extended version of the 
double-folding model (DFM) \cite{Kho16} is used throughout the present work 
to evaluate the real OP for the \cc and \oc systems. The imaginary OP is chosen 
in the phenomenological WS form, with parameters taken from the global 
systematics \cite{Bra97} of the elastic light HI scattering. These WS parameters 
are further fine-tuned to accurately reproduce the diffraction of the elastic 
\oc cross section at forward angles that is determined overwhelmingly by the 
true elastic scattering. In this way, a proper CRC description of the oscillatory 
structure in the elastic \oc cross section at backward angles should allow us 
to properly assess the contribution from the $\alpha$ transfer 
process to the elastic \oc scattering.     

In the first order of the CRC formalism, the coupling between the elastic 
scattering $^{12}$C ($^{16}$O,$^{16}$O)$^{12}$C and direct $\alpha$ transfer 
$^{12}$C ($^{16}$O,$^{12}$C)$^{16}$O channels shown in Fig.~\ref{f1} are taken 
into account explicitly. The strength of the $\alpha$ transfer in the elastic 
\oc scattering is directly proportional to the probability of the dissociation 
of $^{16}$O into the $\alpha$ particle and $^{12}$C nucleus, i.e., the $\alpha$ 
spectroscopic factor $S_\alpha$ \cite{Arima,Sat83,Tho09,Volya15}. The $S_\alpha$ 
values deduced from the earlier CRC \cite{Szi02} and DWBA \cite{Mor11,Hama11} 
calculations, taking into account only the direct $\alpha$ transfer in the elastic 
\oc scattering, are much larger than those predicted by the SM and cluster model 
calculations \cite{Volya15,Yama12}. Such a disagreement clearly indicates that the 
higher-order, indirect $\alpha$ transfer contributions might not be negligible. 
For example, the indirect $\alpha$ transfer via the $2^+_1$ excitation of the 
$^{12}$C core is expected to be significant because of the large $S_\alpha$ values 
predicted by the SM and $\alpha$-cluster model calculations 
\cite{Arima,Volya15,Yama12,Volya17} for the dissociation 
$^{16}$O$_{\rm g.s.}\to\alpha+^{12}$C$_{2^+_1}$. To shed more light on this problem, 
we have performed in the present work a multistep CRC analysis of the elastic 
\oc scattering, taking into account explicitly the contributions from both the 
direct and indirect $\alpha$ transfers (via the excited states of $^{12}$C and $^{16}$O), 
with the real OP and inelastic scattering form factors for the considered excited
states given by the DFM \cite{Kho16,Kho00}. The $S_\alpha$ values predicted recently 
by the large scale SM calculation \cite{Volya15,Volya17} are used consistently 
in the present work. The possible CRC contributions of the nucleon transfer and 
indirect transfer of $^3$He and $^3$H clusters to the enhancement of the elastic 
\oc cross section at backward angles are also studied. 

\section{Optical model analysis of the elastic \oc scattering}
\label{sec2}
Given the difficulty mentioned above in the OM description of the elastic 
\oc scattering at low energies, a proper choice of the complex, energy
dependent OP for this system is a prerequisite for any CC or CRC study of the 
nonelastic processes induced by the \oc collision. While the elastic \oc data 
at backward angles might include the contribution from the $\alpha$ transfer, 
the Fraunhofer-type oscillation observed at forward angles is generated 
entirely by the elastic scattering, and a properly chosen OP for the \oc system 
should reproduce the forward-angle elastic data as accurately as possible. 
With the increasing energy, when more reaction channels are open, the enhancement 
of the elastic \oc cross section at large angles is gradually disappeared, giving 
rise to the exponential fall-off of the elastic cross section caused by a stronger 
absorption \cite{Brau82}. When the energy reaches the range of refractive energies 
of about 10 to 40 MeV/nucleon, the elastic \oc scattering becomes strongly refractive 
and the nuclear rainbow appears. Such data are indispensable in validating the 
prediction of different theoretical methods, like the DFM which derives the real 
\AA OP from the realistic densities of the two colliding nuclei and appropriate 
nucleon-nucleon (NN) interaction between the projectile- and target nucleons. 
The recent (mean-field based) version of the DFM was shown \cite{Kho16} to account 
very well for the energy dependence of the real OP, and the (energy dependent) 
folded potential $U_{\rm F}$ \cite{Kho16} obtained with the CDM3Y3 density dependent 
NN interaction \cite{Kho97} is used in the present OM analysis of the elastic \oc 
scattering as the real OP.  
\begin{figure}\vspace*{-2cm}\hspace*{0cm}
\includegraphics[width=0.9\textwidth]{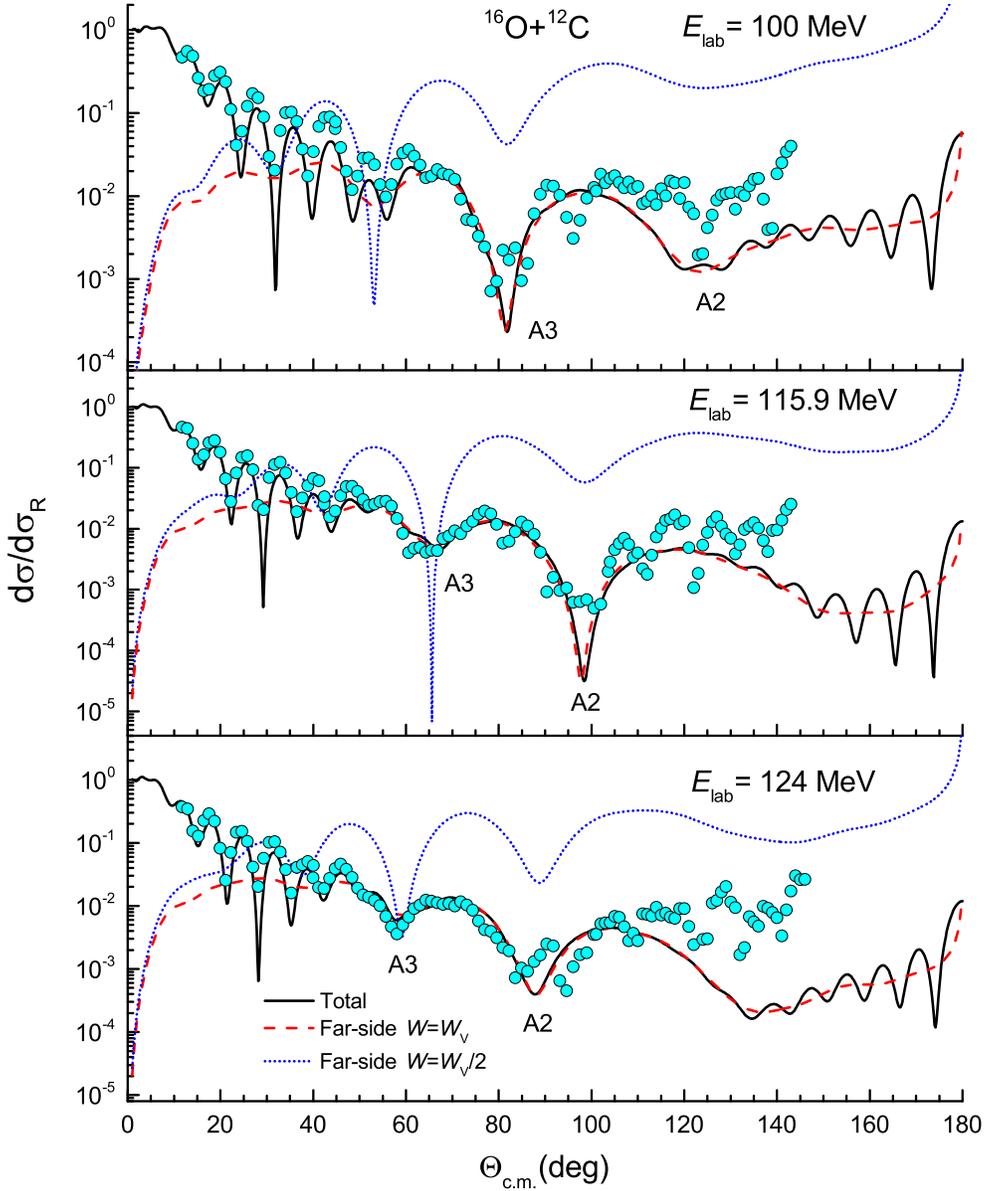}\vspace*{-2cm}
 \caption{OM description of the elastic \oc data at $E_{\rm lab}=100$, 
115.9, and 124 MeV \cite{Nico00} given by the best-fit real folded OP 
and WS imaginary OP (see parameters in Table~\ref{tOP}). The far-side scattering 
cross sections are given by the near-far decomposition (\ref{NFdec}) using 
the same real folded OP but with different strengths $W_V$ of the WS imaginary 
OP (dashed and dotted lines). Ak is the k-th order Airy minimum.} 
\label{fOM1}
\end{figure}
The imaginary (absorptive) OP is due to the open nonelastic channels, and is 
usually assumed in the standard WS form. Thus, the total OP at the 
internuclear distance $R$ is determined as
\begin{equation}
 U(R)=N_RU_{\rm F}(E,R)-\frac{iW_V }{1+\exp[(R-R_V)/a_V]}+V_{\rm C}(R).
 \label{eOP} 
\end{equation}
The Coulomb potential $V_{\rm C}(R)$ is obtained by folding two uniform 
charge distributions \cite{Pol76}, chosen to have RMS charge radii 
$R_C=3.17$ and 3.54 fm for $^{12}$C and $^{16}$O, respectively. Such a choice 
of the Coulomb potential was shown to be accurate up to small radii where 
the nuclear interaction becomes dominant \cite{Bra97}. The ground state (g.s.) 
densities of $^{16}$O and $^{12}$C used in the DFM calculation were taken as 
Fermi distributions with parameters \cite{Kho01} chosen to reproduce the empirical 
matter radii of these nuclei. The OM calculations were made using the code 
ECIS97 written by Raynal \cite{Raynal}. The renormalization factor $N_R$ of the 
real folded OP and parameters of the WS imaginary OP were adjusted by the best 
OM description of the elastic data, especially, the data points at the most 
forward angles. 
\begin{table}[bht]
\caption{The best-fit OP parameters (\ref{eOP}) used in the OM analysis of the
elastic \oc scattering at $E_{\rm lab}=100-300$ MeV. $J_R$ and $J_W$ are the 
volume integrals (per interacting nucleon pair) of the real and imaginary parts 
of the OP, respectively.} \label{tOP} 
\begin{center}
 \begin{tabular}{|c|c|c|c|c|c|c|c|c|} \hline
$E_{\rm lab}$ & $N_R$ & $J_R$ & $W_V$ & $R_V$ & $a_V$ & $J_W$ & $\sigma_R$ & Data \\
		(MeV) &  & (MeV~fm$^3$) & (MeV) & (fm) & (fm) & (MeV~fm$^3$) & (mb) &  \\ \hline
		100   & 1.006 & 332.1 & 11.21 &	6.020 &	0.52  &	57.0  &	1401 & \cite{Nico00} \\ 
		115.9 & 1.002 & 328.8 &	13.70 & 5.781 &	0.60  &	63.9  &	1464 & \cite{Nico00} \\ 
		124   & 1.000 & 327.1 &	14.50 &	5.800 &	0.60  &	68.2  &	1485 & \cite{Nico00} \\		
 132   & 1.020 & 332.6 &	14.00 &	5.853 &	0.72  &	70.4  &	1638 & \cite{Oglo98,Oglo00} \\	
		300   & 0.960 & 293.6 &	26.37 &	5.632 &	0.68  &	117.5 &	1655 & \cite{Bra01} \\ \hline
 \end{tabular}
\end{center}
\end{table}

Very helpful for the illustration of the refractive structure of the nuclear 
rainbow is the near-far decomposition of the elastic scattering amplitude based on 
the method developed by Fuller \cite{Ful75}. Namely, by decomposing the Legendre 
function $P_l(\cos\theta)$ into waves traveling in $\theta$ that are running in 
the opposite directions around the scattering center, the elastic amplitude $f(\theta)$ 
can be expressed in terms of the near-side ($f_{\rm N}$) and far-side ($f_{\rm F}$) 
components as
\begin{equation}
  f(\theta)=f_{\rm N}(\theta)+f_{\rm F}(\theta)=\frac{i}{2k}\sum_l (2l+1)
  A_l\left[\tilde Q_l^{(-)}(\cos\theta)+\tilde Q_l^{(+)}(\cos\theta)\right],\
 \label{NFdec}
 \end{equation}
 \begin{equation}
 {\rm where}\ \ \tilde Q_l^{(\mp)}(\cos\theta)={1\over 2}
  \left[P_l(\cos\theta)\pm {2i\over\pi}Q_l(\cos\theta)\right], \nonumber
  \end{equation}
and $Q_l(\cos\theta)$ is the Legendre function of the second kind. The
amplitude $f_{\rm N}(\theta)$ represents the waves deflected to the direction 
of $\theta$ on the near side of the scattering center, and the waves 
traveling on the opposite, far side of the scattering center to the same angle 
$\theta$ give rise to the far-side amplitude $f_{\rm F}(\theta)$. Therefore, 
the near-side scattering occurs mainly at the surface, while the far-side 
(refractive) scattering penetrates more into the interior of the \AA system. 
The broad oscillation of the far-side cross section is directly associated 
with the Airy structure of the nuclear rainbow \cite{Kho07r,Bra96}.

\begin{figure}\vspace*{-1cm}\hspace*{0cm}
\includegraphics[width=1.0\textwidth]{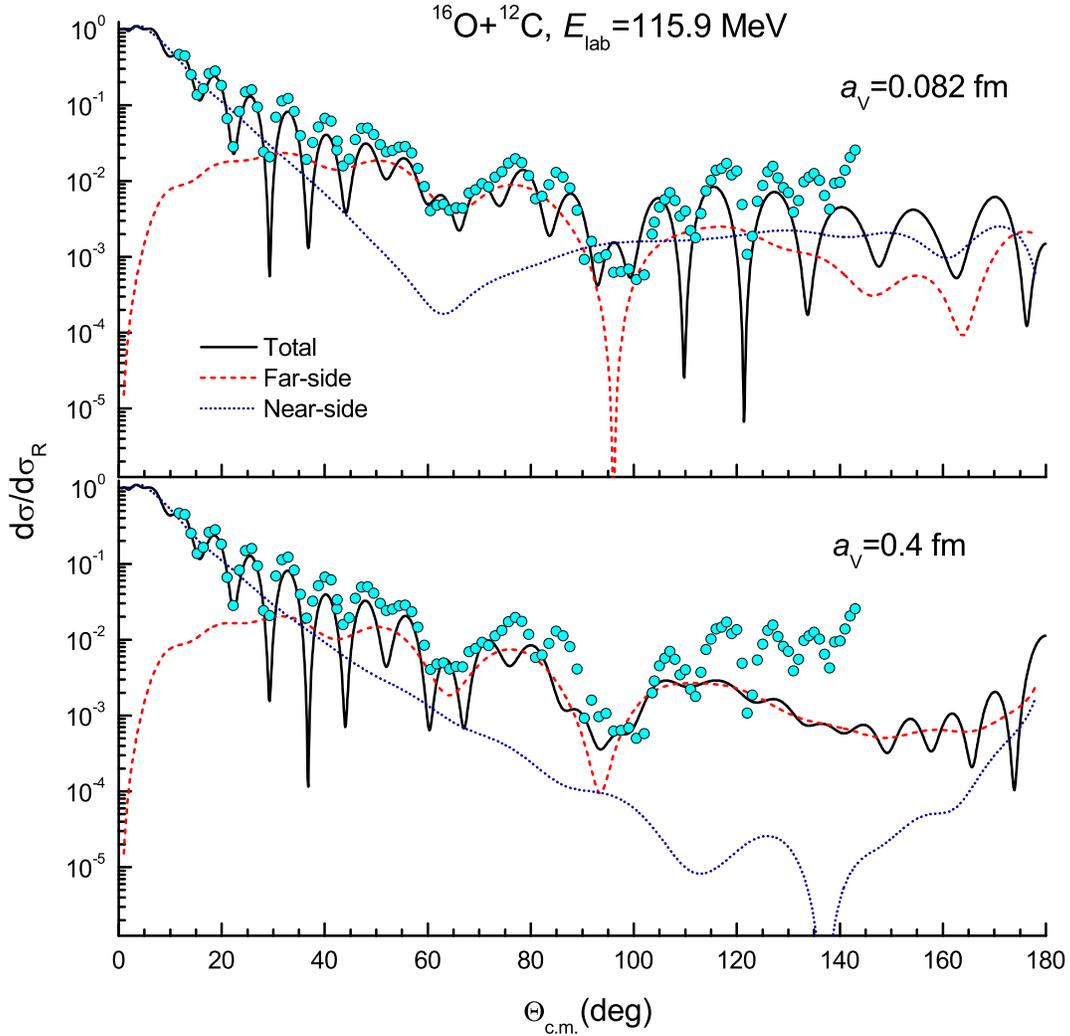}\vspace*{-1.5cm}
 \caption{OM description of the elastic \oc data at $E_{\rm lab}=115.9$ MeV 
\cite{Nico00} given by the best-fit real folded OP and WS imaginary OP containing 
both the volume and surface terms (solid lines). The different behaviors of the 
near-side cross section at large angles are caused by two different values of the 
diffuseness $a_V$ of the volume WS term of the imaginary OP.}
\label{f116a}
\end{figure}
The OM description of the elastic \oc scattering data at $E_{\rm lab}=100-124$ MeV 
is shown in Fig.~\ref{fOM1}. The realistic OP family obtained in the folding model 
analysis \cite{Kho16} of the elastic \oc scattering at the refractive energies 
was extrapolated to lower energies, and the best OM fit to the elastic data from 
the forward angles up to the center-of-mass (c.m.) angles around $100^\circ$ 
has been achieved with the real folded OP renormalized by the factor $N_R$ close 
to unity (see Table~\ref{tOP}). The obtained volume integrals (per interacting 
nucleon pair) of the real ($J_R$) and imaginary ($J_W$) potentials agree well 
with the global systematics of the elastic scattering of light HI given by both 
the OM analysis of the elastic data and prediction by the dispersion relation 
that links the real and imaginary parts of the OP (see, e.g., Fig.~6.7 in 
Ref.~\cite{Bra97}). Because the refractive Airy structure determined by the 
far-side scattering is frequently weakened by the absorption, the OM calculation 
was done also with a strength of the imaginary WS potential reduced by 50\%, 
and the results (see the far-side cross sections shown in Fig.~\ref{fOM1}) 
reproduce nicely the broad Airy oscillation pattern established earlier 
in the OM analyses of the elastic \oc data over a wide range of energies 
\cite{Kho16,Oglo00}. At low energies ($E_{\rm lab}\lesssim 132$ MeV), the first 
Airy minimum A1 is located in the backward region and strongly distorted by 
a quickly oscillating cross section that is likely due to the elastic $\alpha$ 
transfer process \cite{Brau82,vOe75}. 

In the OM description of the elastic \oc scattering, the quick Fraunhofer oscillation 
of the elastic cross section at the forward angles is well known to be due to the 
interference between the near-side and far-side scattering (\ref{NFdec}). 
Because the near-side scattering contributes mainly to the diffraction at forward 
angles, the near-far interference is usually weak at medium and large angles where 
only the far-side scattering amplitude survives \cite{Kho07r}. However, the quickly 
oscillating cross section at large angles observed in the elastic \oc scattering at low 
energies (see Fig.~\ref{fOM1}) seems to result from some interference pattern 
that distorts the smooth far-side cross section there. The original OM analysis 
of these data by Nicoli {\it et al.} \cite{Nico00} based on different choices 
of the OP, including the model-independent spline shape for the real OP, 
has shown that the observed oscillatory enhancement of the elastic cross section 
at backward angles can be reproduced only with a very \emph{small diffuseness} 
of the WS imaginary OP. 

To illustrate this effect, we have performed the OM calculation using the same 
best-fit real folded OP and WS imaginary OP containing both the volume and surface
terms as that used in Ref.~\cite{Nico00}, with parameters adjusted by the best 
OM fit to the data. The best-fit WS parameters turned out to be quite close to those 
obtained in Ref.~\cite{Nico00}, with a very small diffuseness $a_V\approx 0.08$ fm 
of the volume WS term. From the OM results shown in the upper panel of Fig.~\ref{f116a}
one can see that the main impact by such a small diffuseness of the WS term is the
unusually enhanced strength of the near-side scattering at backward angles, and the  
near-far interference gives rise then to the enhanced oscillation of the elastic 
cross section. Keeping the same WS imaginary OP but with a larger $a_V\approx 0.4$ fm, 
the near-side scattering is weakened substantially and the OM calculation fails 
again to describe the oscillating cross section at large angles (see lower panel 
of Fig.~\ref{f116a}). In general, the near-side (surface) scattering occurs mainly 
at forward angles, and to boost the strength of the near-side scattering at backward 
angles using the WS absorptive potential with an extremely small diffuseness is just 
a computational technique to reproduce the enhanced oscillation of the elastic \oc 
cross section at backward angles, without any physics explanation of that phenomenon.  

While a widely expected physics scenario in this case is the strong contribution 
from the $\alpha$ transfer channels to the elastic \oc scattering \cite{Brau82,vOe75},
Ohkubo and Hirabayashi \cite{Ohku14-2} have proposed a different scenario in their 
recent CC study of the elastic \oc scattering at $E_{\rm lab}=115.9$ MeV, where the 
oscillatory pattern at large angles was interpreted as a result of the interference 
between the elastic scattering wave and ``external reflective" wave (caused by 
the nuclear excitations taken into account in the CC calculation). A closer look 
at these results finds that such an external reflective wave is caused also by a 
small diffuseness ($a_V\approx 0.2$ fm) of the volume WS imaginary potential used 
in the CC calculation, because the same CC calculation using a larger diffuseness 
$a_V\approx 0.4$ fm (see Fig.~3 in Ref.~\cite{Ohku14-2}) fails again to describe
the large-angle oscillation of the elastic cross section, in a manner
similar to that shown in the lower panel of Fig~\ref{f116a}. 

\section{CRC study of the elastic alpha transfer 
 $^{12}$C ($^{16}$O,$^{12}$C)$^{16}$O}
It is well established that the elastic scattering of nearly identical nuclei 
at low energies often involves the elastic transfer \cite{Frahn84}, which 
gives rise to a quickly oscillating elastic cross section at backward angles. 
For the \oc system, the elastic $\alpha$ transfer from $^{16}$O to $^{12}$C 
leads to the final state that is indistinguishable from that of the true elastic 
\oc scattering (see Fig.~\ref{f1}). Therefore, the total elastic amplitude 
should be a coherent sum of the elastic scattering amplitude $f_{\rm ES}$ 
and elastic $\alpha$ transfer amplitude $f_{\rm ET}$. The interference between 
$f_{\rm ES}$ and $f_{\rm ET}$ gives rise to the oscillatory elastic cross section 
at large angles, which is essentially the oscillation due to the symmetry of the 
two identical $^{12}$C cores \cite{Frahn84,vOe75}. Given the significant $\alpha$ 
spectroscopic factors predicted by the large scale SM calculation \cite{Volya15} 
for different paths of the dissociation of $^{16}$O into $\alpha$-particle and 
the $^{12}$C core, the coupled channel contribution from the direct and indirect 
$\alpha$ transfers to the total elastic \oc cross section should be significant at low 
energies. Therefore, we have performed in the present work a detailed CRC analysis 
of the elastic \oc scattering at the energies $E_{\rm lab}=100-300$ MeV, taking 
into account explicitly the couplings between the elastic scattering channel and 
different $\alpha$ transfer channels using the code FRESCO written by 
Thompson \cite{Tho88}. 

\subsection{CRC formalism} 
\label{sec3a}
We give here a brief description of the multichannel CRC method to study the 
contributions of different $\alpha$ transfer channels to the elastic \oc 
scattering. In general, the coupled equation in the post form 
for a particular channel $\beta$ can be written as
\begin{equation}
(E_\beta-T_\beta-U_\beta)\chi_\beta=\sum_{\beta'\neq\beta,x=x'}
\langle\beta|V|\beta'\rangle\chi_{\beta'}+\sum_{\beta'\neq\beta,x\neq x'}
[\langle\beta|W_{\beta'}|\beta'\rangle +\langle\beta|\beta'\rangle
(T_{\beta'}+U_{\beta'}-E_{\beta'})]\chi_{\beta'}, \label{crc1}
\end{equation}
where $\beta'$ is the scattering or transfer channel different from $\beta$, 
$x$ and $x'$ are the partitions associated with the considered transfer 
process. $\chi_\beta$ and $\chi_{\beta'}$ are the relative-motion wave functions 
while $U_\beta$ and $U_{\beta'}$ are the (diagonal) optical potentials in these 
two channels. Due to the identity of the entrance and exit channels, all post form 
formulas are equivalent to the prior ones, and the transfer interactions $W_\beta$ 
can be determined \cite{Sat83,Tho09} as 
\begin{equation}
 W_\beta= V_{\alpha+^{12}\text{C}}+(U_{^{12}\text{C}+^{12}\text{C}}-
 U_{^{16}\text{O}+^{12}\text{C}}), \label{crc2}
\end{equation}
where $(U_{^{12}\text{C}+^{12}\text{C}}-U_{^{16}\text{O}+^{12}\text{C}})$ 
is the complex remnant term which is the difference between the core-core OP and 
that of the exit channel. $V_{\alpha+^{12}\text{C}}$ is the binding potential 
of the $\alpha$ cluster inside the $^{16}$O nucleus, which has been assumed in 
the standard WS form in our CRC calculation. The CRC equations  
(\ref{crc1})-(\ref{crc2}) are solved iteratively, with the finite-range complex 
remnant terms and non-orthogonality corrections properly taken into account 
\cite{Tho88}. The diagonal OP for the considered channels are determined by
Eq.~(\ref{eOP}) using the real double-folded potential \cite{Kho16} and WS 
imaginary potential. The (complex) inelastic scattering form factors for the 
nuclear transitions considered in the CRC study are also obtained in the DFM 
calculation \cite{Kho00} using the same CDM3Y3 interaction \cite{Kho16}, with 
a complex density dependence suggested in Ref.~\cite{Kho08}. 
For the $\alpha$ transfer, the internal state of the bound $\alpha$ cluster 
is assumed to be $1s$ state. Then, the relative-motion wave function 
$\Phi_{NL}(\bm{r}_{\alpha+{\rm ^{12}C}})$ of the $\alpha$+$^{12}$C system 
($L$-wave state) has the number of radial nodes $N$ determined by the Wildermuth 
condition \cite{Sat83,Tho09}, so that the total number of the oscillator quanta 
$\mathcal{N}$ is conserved  
\begin{equation}
 \mathcal{N}=2(N-1)+L=\sum_{i=1}^{4}2(n_i-1)+l_i, \label{crc3}
\end{equation}
where $n_i$ and $l_i$ are the principal quantum number and orbital momentum of each 
constituent nucleon in the $\alpha$ cluster. Using the fixed WS geometry  
($R=4.148$ fm and $a=0.55$ fm in the 2-channel CRC calculation of the direct 
$\alpha$ transfer; $R=3.683$ fm and $a=0.55$ fm in the multi-channel
CRC calculation of the direct and indirect $\alpha$ transfer) the wave function 
$\Phi_{NL}(\bm{r}_{\alpha+{\rm ^{12}C}})$ is obtained with the depth of the WS 
potential adjusted to reproduce the $\alpha$ separation energy 
$E_\alpha$ given by the relation
\begin{equation}
E_\alpha(J^\pi_i) = E_\alpha({\rm g.s.}) - E(^{16}{\rm O}^*)+E(^{12}{\rm C}^*),
\label{crc4}
\end{equation}
where the $\alpha$ separation energy of $^{16}$O in the ground state is 
$E_\alpha({\rm g.s.})=7.162$ MeV \cite{Til93}, $E(^{16}{\rm O}^*)$ and $E(^{12}{\rm C}^*)$ 
are the excitation energies of $^{16}$O and the $^{12}$C core, respectively. 
The solutions $\chi_\beta$ of the CRC equations (\ref{crc1})-(\ref{crc3}) are used 
to determine the elastic scattering $f_{\rm ES}$ and $\alpha$ transfer $f_{\rm ET}$ 
amplitudes. The total elastic \oc cross section is given \cite{Tho88} by
\begin{equation}
\frac{d\sigma(\theta)}{d\Omega}=\left|f(\theta)\right|^2=\left|f_{\rm ES}(\theta)
 +f_{\rm ET}(\pi-\theta)\right|^2, \label{fsig}
\end{equation}  
where the elastic $\alpha$ transfer amplitude at the c.m. angle ($\pi-\theta$) 
is coherently added to the elastic scattering amplitude at the c.m. angle 
$\theta$ (as illustrated in Fig.~\ref{f1}).

The CRC calculation requires the input of the spectroscopic amplitude 
$A_{NL}$ \cite{Arima} which is used to construct the dinuclear overlap as 
\begin{equation}
\langle{\rm ^{12}C}|^{16}{\rm O}\rangle=A_{NL}(^{16}{\rm O},{\rm ^{12}C})
\Phi_{NL}(\bm{r}_{\alpha+{\rm ^{12}C}}). \label{Amp0}
\end{equation}     
The $\alpha$ spectroscopic factor is then determined as $S_\alpha=|A_{NL}|^2$. 
In the present CRC analysis, we have used $S_\alpha$ predicted recently by the 
large scale SM calculation, the so-called cluster-nucleon configuration interaction 
model \cite{Volya15,Volya17}. In this approach, $S_\alpha$ of $^{16}$O 
has been obtained in the unrestricted single-nucleon \emph{p-sd} configuration 
space, using the realistic SM Hamiltonian and new definition of $S_\alpha$. 
The norm kernel originating from the full antisymmetrization and orthonormalization 
of the multinucleon clustering wave functions was found to be very substantial, 
which increases $S_\alpha$ for $^{16}$O in the g.s. from around 0.29 as given by the 
traditional definition of $S_\alpha$ (see Table I of Ref.~\cite{Kra17} and 
discussion thereafter) to 0.794 (see Table III of Ref.~\cite{Volya15}). In fact, 
the use of $S_\alpha$ determined in the traditional SM method was questioned some 
40 years ago by Fliessbach \cite{Flis76,Flis77}, and a new definition of the cluster 
spectroscopic factor similar to that adopted in Ref.~\cite{Volya15} was used in 
the microscopic cluster decay study \cite{Lovas,Betan}. Although it was shown years 
ago \cite{Lov85} that the use of the new definition of $S_\alpha$ in a transfer 
reaction calculation does not require any reformulation of the DWBA or CRC formalism, 
the present CRC calculation seems to be the first attempt to use the newly defined 
$S_\alpha$ in the study of the $\alpha$ transfer reaction. 

\subsection{Direct $\alpha$ transfer} 
\label{sec3b}
The observed oscillatory enhancement of the elastic \oc cross section at large 
angles was repeatedly discussed in the past as due to the contribution from the 
elastic $\alpha$ transfer \cite{Brau82,vOe75,Frahn84}. A straightforward method 
to estimate the strength of the direct (elastic) $\alpha$ transfer in the elastic 
\oc scattering is to add the elastic $\alpha$ transfer amplitude in the DWBA 
to the elastic scattering amplitude \cite{Mor11,Hama11}. A more consistent approach
is to solve the CRC equations (\ref{crc1})-(\ref{crc4}), coupling explicitly the 
elastic scattering and direct $\alpha$ transfer channels \cite{Szi02}.       
\begin{figure}\vspace*{-1cm}\hspace*{0cm}
\includegraphics[width=0.9\textwidth]{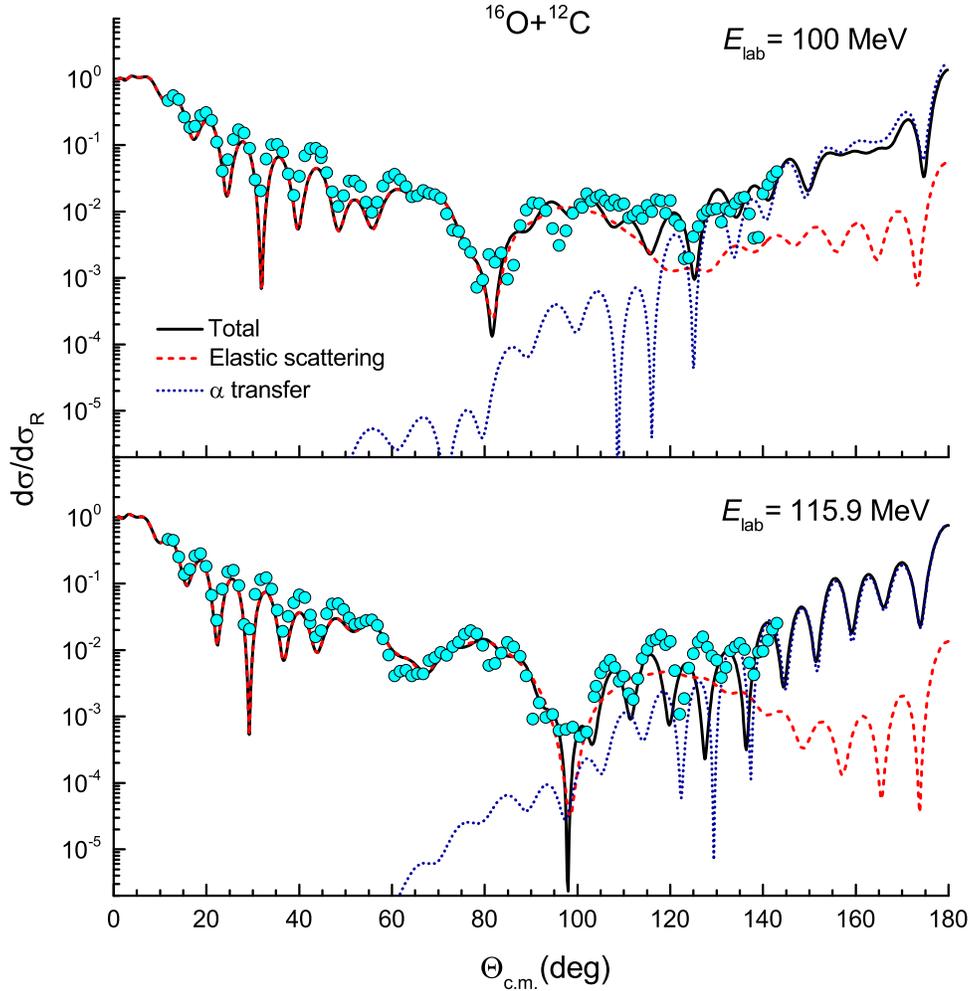}\vspace*{-1cm}
 \caption{CRC description (solid lines) of the elastic \oc data measured 
at $E_{\rm lab}=100$ and 115.9 MeV \cite{Nico00}. The true elastic scattering 
cross section (dashed lines) are obtained with the real folded 
potential and WS imaginary potential taken from Table~\ref{tOP}. The elastic 
$\alpha$ transfer cross section (dotted lines) are given by the best-fit 
$\alpha$ spectroscopic factor $S_\alpha\approx 1.96$.} 
\label{fET1}
\end{figure}
\begin{figure}\vspace*{-1cm}\hspace*{0cm}
\includegraphics[width=0.9\textwidth]{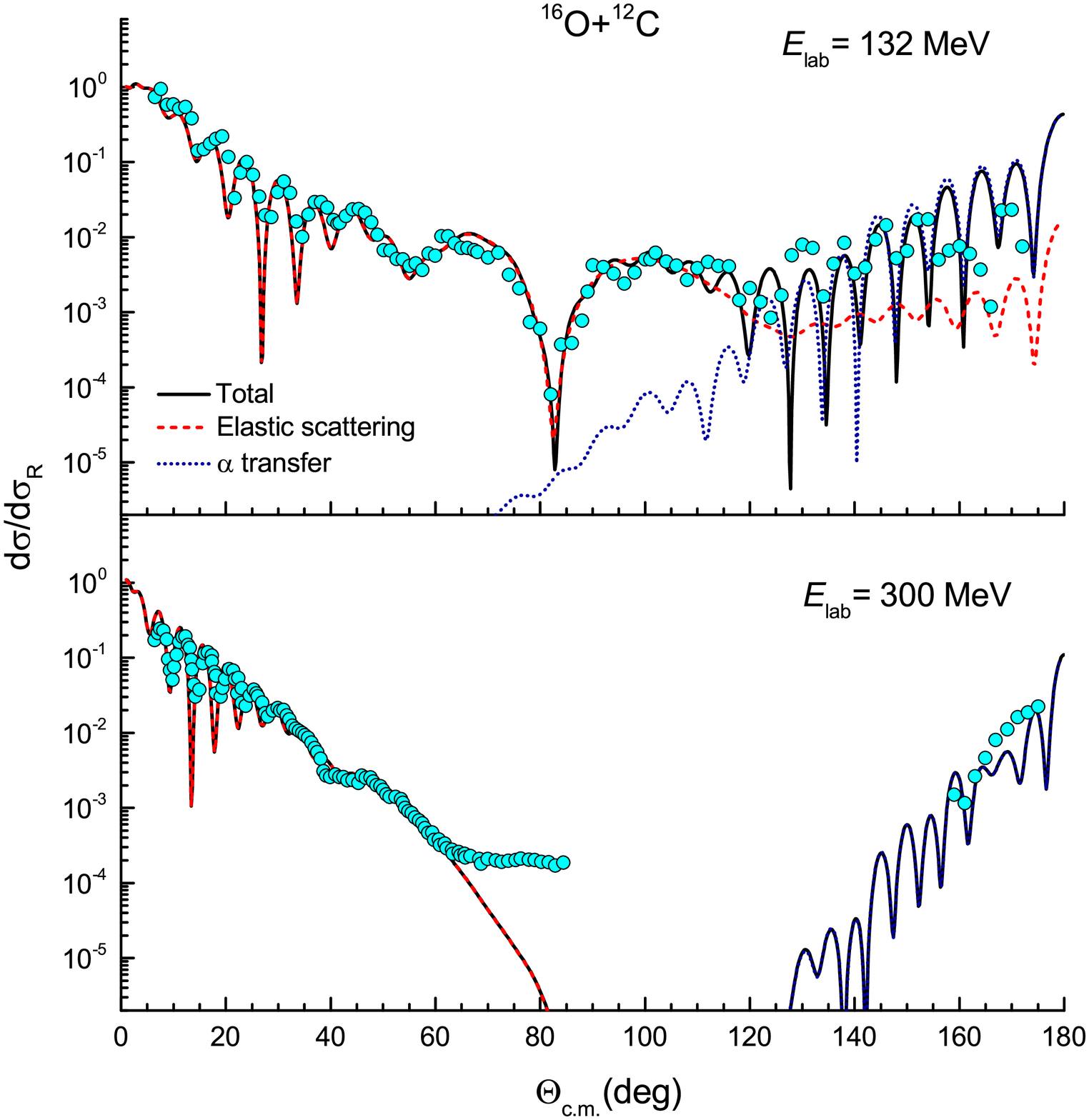}\vspace*{-1cm}
 \caption{The same as Fig.~\ref{fET1} but for the elastic \oc data measured 
at $E_{\rm lab}=132$ \cite{Oglo98,Oglo00}, and 300 MeV \cite{Bra01}.}. 
\label{fET2}
\end{figure}
We discuss briefly here the results of our CRC calculation of the direct
$\alpha$ transfer using the (real folded + imaginary WS) optical potentials
(\ref{eOP}) for the \oc and \cc systems. Assuming the $\alpha$-clustering of the 
$p$-shell nucleons in $^{16}$O, the direct elastic $\alpha$ transfer proceeds via 
the $S$ state ($L=0$) of the $\alpha$+$^{12}$C system, which implies $N=3$ 
in the Wildermuth's rule (\ref{crc3}) including the origin and excluding 
the infinity. In this case, the WS depth of the $\alpha$ binding potential was 
adjusted to reproduce the $\alpha$ separation energy $E_\alpha({\rm g.s.})\approx 
7.162$ MeV. The total elastic \oc cross sections given by the present CRC 
calculation are compared with the elastic \oc data measured at $E_{\rm lab}=100-300$ 
MeV \cite{Nico00,Oglo98,Oglo00,Bra01} in Figs.~\ref{fET1} and \ref{fET2}. The use
of the $\alpha$ spectroscopic factor predicted by the SM \cite{Volya15} gives a  
very weak elastic transfer cross section that cannot account for the enhanced
oscillating cross section at backward angles. It is an indication of the
strong contribution from the indirect $\alpha$ transfer to the elastic \oc cross
section. A simple (effective) way of the DWBA or CRC analysis of the direct transfer
reaction is to treat the cluster spectroscopic factor as a free parameter to 
be adjusted to the best fit of the transfer data. In such an approach, our CRC
calculation gives consistently a good description of the elastic \oc data at
different energies using the same $\alpha$ spectroscopic factor 
$S_\alpha\approx 1.96$ for the g.s. of $^{16}$O (see Figs.~\ref{fET1} and 
\ref{fET2}). The explicit comparison of the true elastic \oc scattering and direct 
$\alpha$ transfer shows that the oscillatory enhancement of the elastic cross 
section at backward angles is due to the elastic $\alpha$ transfer \Atrans process. 
We also found that the back coupling from the direct $\alpha$ transfer to the 
elastic \oc scattering at forward angles is not significant, and the same complex 
OP's as those used in the OM calculation discussed in Section~\ref{sec2} 
(see Table~\ref{tOP}) can be used in the CRC calculation. We note that the data 
points at the most backward angles measured at $E_{\rm lab}=300$ MeV \cite{Bra01} 
are described by the present CRC result as due entirely to the elastic $\alpha$ 
transfer. We have considered these data points in the OM analysis of the elastic 
\oc scattering at 300 MeV, and they cannot be described by the standard OM 
calculation, despite a broad variation of the strength and shape of the complex OP. 
Consequently, the 300 MeV data points at the most backward angles can be used to 
gauge the strength of the $\alpha$ spectroscopic factor in the CRC analysis 
of the elastic \oc scattering. The best-fit $S_\alpha\approx 1.96$ obtained in the 
2-channel CRC analysis agrees reasonably with that deduced earlier (see Table~\ref{tSa}) 
from the DWBA and CRC studies of the elastic \oc scattering at low energies.
\begin{table}[bht]
\caption{$\alpha$ spectroscopic factor $S_\alpha$ deduced from the 2-channel CRC 
analysis of the elastic scattering and direct $\alpha$ transfer in the \oc system, 
in comparison with that deduced earlier from the DWBA and CRC studies of the elastic 
$\alpha$ transfer reaction $^{12}$C ($^{16}$O,$^{12}$C)$^{16}$O.} 
\label{tSa}\vspace{0.5cm}
 \begin{tabular}{|c|c|c|} \hline 
	$E_\text{lab}$ (MeV) & $S_\alpha$ & Reference \\ \hline  
   $100-300$ & $1.96$ & Present work \\
	 $20-35$   & $1.0-1.96$ & \cite{vOe75} \\ 
	 $100-124$ & $1.21-1.96$ & \cite{Szi02} \\ 
	 $20-132$  & $1.45-1.58$ & \cite{Mor11} \\ 
	 $132,181$ & $0.49-0.81$ & \cite{Grid13} \\ 
	 $28-61.5$ & $1.59-3.00$ & \cite{Hama14} \\ \hline 		
 \end{tabular}
\end{table}
However, most of the empirical values of the $\alpha$ spectroscopic factor 
$S_\alpha$ shown in Table~\ref{tSa} seem to be much larger than that predicted 
by the recent large scale SM calculation ($S_\alpha\approx 0.8$) \cite{Volya15} 
or 4$\alpha$ cluster model ($S_\alpha\approx 0.6$) \cite{Yama12}.

\subsection{Indirect $\alpha$ transfer via the excited states of $^{16}$O and the 
 $^{12}$C core} \label{sec3c}
\begin{figure}\vspace*{0cm}\hspace*{0cm}
\includegraphics[width=0.9\textwidth]{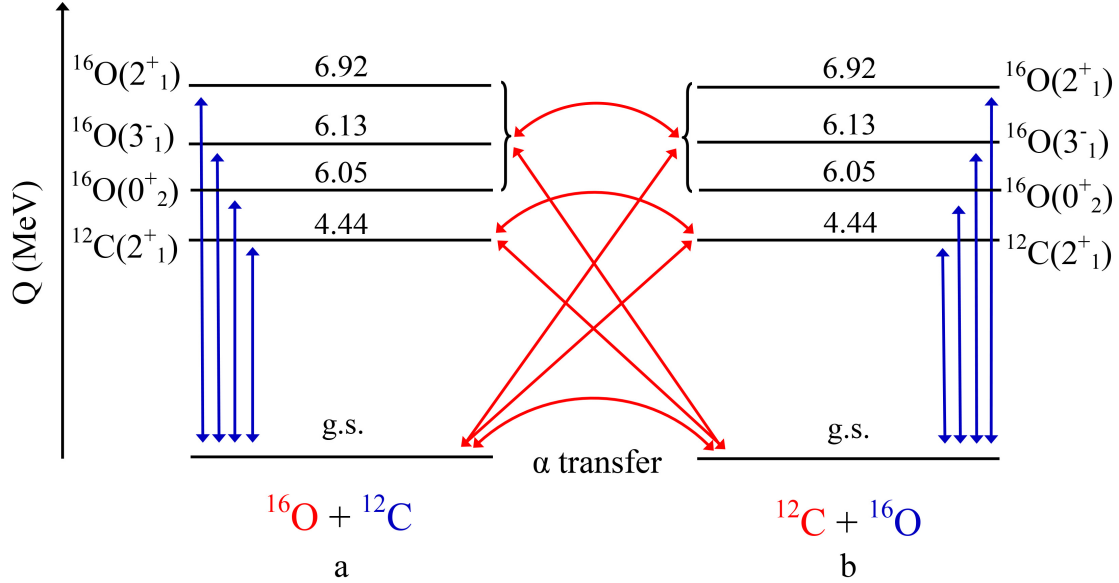}\vspace*{0cm}
 \caption{Coupling scheme of the 10 reaction channels taken into account 
in the present CRC analysis of the elastic \oc scattering that includes both the 
direct and indirect $\alpha$ transfer processes (see details in text).} \label{fCoup}
\end{figure}

The scenario of the direct $\alpha$ transfer presented in Sec.~\ref{sec3b} is sound 
but the question remains about the obtained $\alpha$ spectroscopic factor 
$S_\alpha\approx 1.96$ that is more than twice that predicted by the
SM calculation \cite{Volya15} or 4$\alpha$ cluster model of $^{16}$O \cite{Yama12}. 
Obviously, the various indirect $\alpha$ transfer channels via the excited states 
of $^{16}$O and the $^{12}$C core should also contribute to the total elastic $\alpha$
transfer. For example, the elastic $\alpha$ transfer in the \oc system was predicted 
by the cluster models \cite{Arima,Yama12} to proceed also indirectly through 
the $2^+_1$ excited state of the $^{12}$C core. In general, it is necessary to consider 
the indirect $\alpha$ transfer in a sufficiently large model space consisting of the 
most important excited states of $^{16}$O \cite{Rud10}. The $0^+_2$ (6.05 MeV) and 
$2^+_1$ (6.92 MeV) states of $^{16}$O are the low-lying members of the $K^\pi=0^+$ 
rotational band with a well-developed \ac cluster structure \cite{Enyo17}, and they 
are expected to have a non-negligible contribution to the $\alpha$ transfer in the 
elastic \oc scattering. 
The $3^-_1$ (6.13 MeV) state has a SM-type structure similar to the g.s. of $^{16}$O,
but it can have a strong coupling effect caused by a large octupole transition strength. 
The $0^+_2$, $3^-_1$, $2^+_1$ excited states of $^{16}$O were also predicted by the 
large scale SM \cite{Volya15} to have the $\alpha$ spectroscopic factor 
$S_\alpha\approx 0.535,$ 0.663, and 0.5, respectively, which are comparable with 
$S_\alpha\approx 0.794$ predicted for the g.s. of $^{16}$O. The other excited 
states of $^{16}$O are at the higher energies, with the $\alpha$ spectroscopic 
factors (see Table III in Ref.~\cite{Volya15}) predicted to be much smaller than
those mentioned above. Therefore, we have tried in the present multichannel CRC 
analysis of the elastic \oc scattering to explore the impact of the indirect $\alpha$ 
transfer through the $0^+_2$, $3^-_1$, and $2^+_1$ excited states of $^{16}$O as well 
as the $2^+_1$ state of the $^{12}$C core. The explicit coupling scheme of the 10 
reaction channels considered in the present work is shown in Fig.~\ref{fCoup}.    

To study the indirect $\alpha$ transfers via the excited states of $^{16}$O in the
CRC analysis with the coupling scheme shown in Fig.~\ref{fCoup}, the dinuclear 
overlap is determined by the same relation (\ref{Amp0}), where the orbital momentum 
$L$ of the relative-motion wave function $\Phi_{NL}(\bm{r}_{\alpha+{\rm ^{12}C}})$ 
is assumed to be equal the spin of the excited state of $^{16}$O. For a consistent 
test of the $\alpha$ spectroscopic factors predicted recently in the large scale SM 
calculation by Volya and Tchuvilsky \cite{Volya15,Volya17} that treats the cluster 
channel wave function in a translationally invariant manner \cite{Kra17}, we have used 
these $S_\alpha$ values as fixed parameters in our CRC analysis of the elastic \oc 
scattering. The summary of the CRC inputs for the g.s. and 
excited states of $^{16}$O is given in Table~\ref{tSO}.  
\begin{table}[bht]
\caption{$S_\alpha$ values used in our CRC analysis of the direct and indirect 
$\alpha$ transfer via the g.s. and excited states of $^{16}$O were taken from the 
results of the large scale SM calculation \cite{Volya15}. The number of radial 
nodes $N$ and orbital momentum $L$ of the corresponding $\alpha$+$^{12}$C configurations 
are given by the Wildermuth rule (\ref{crc3}).} 	\label{tSO} \vspace{0.5cm}
		\begin{tabular}{|c|c|c|c|c|c|} \hline
	$\ J^\pi\ $ &   $\ E_x\ $  &  $\ N\ $  & $\ L\ $ & $\ S_\alpha\ $ \\
			          &    (MeV) &     &     &            \\ \hline
			 $0^+_1$  &     0.000 &   3   &  0  &   0.794    \\
			 $0^+_2$  &     6.049 &   5   &  0  &   0.535    \\
			 $3^-_1$  &     6.130 &   2   &  3  &   0.663    \\
			 $2^+_1$  &     6.917 &   4   &  2  &   0.500 \\ \hline
		\end{tabular}
\end{table} 
The complex inelastic $^{16}$O$_{\rm g.s.}\to ^{16}$O$^*$ form factors were calculated 
in the generalized DFM \cite{Kho00}, using the transition densities of the $0^+_2$, 
$3^-_1$, and $2^+_1$ states obtained in the Orthogonality Condition Model (OCM)  
by Okabe \cite{Okabe} and the complex CDM3Y3 density dependent interaction \cite{Kho16}. 
The OCM transition densities were slightly renormalized to reproduce the measured 
transition strengths \cite{Miska}, M(E0) = $3.55\pm 0.21\ e~{\rm fm}^{2}$, 
B(E3) = $1490 \pm 70\ e^2 {\rm fm}^{6}$, and B(E2) = $39.3 \pm 1.6\ e^2{\rm fm}^{4}$
of the $0^+_2$, $3^-_1$, and $2^+_1$ states, respectively. 

The dissociation of $^{16}$O into the $\alpha+^{12}$C$_{2^+_1}$
configuration has been discussed earlier in the SM and nuclear cluster studies
\cite{Arima,Yama12,Rot68,Su76,Ku73}. Given $^{16}$O in its g.s., we need to 
input $N=2$ and $L=2$ into the Wildermuth's rule (\ref{crc3}) 
for the $\alpha+^{12}$C$_{2^+_1}$ configuration. According to the SM studies 
\cite{Arima,Volya15,Rot68,Ku73,Volya17} the 2$^+_1$ and g.s. of $^{12}$C are  
of the same SU(3) shell structure ($\alpha$ composed of the same clustering nucleons 
coupled to the $^{12}$C core in its g.s. or an excited states) and should have, 
therefore, the same $\alpha$ spectroscopic factor. However, this $S_\alpha$ value 
must be enhanced by the number of the possible $M$-substates of the 
$\alpha+^{12}$C$_{2^+_1}$ configuration \cite{Arima}, so that 
\begin{equation}
A^2_{NL}(^{16}{\rm O}_{\rm g.s.},{\rm ^{12}C}^*(L))=(2L+1)
A^2_{N=3,L=0}(^{16}{\rm O}_{\rm g.s.},{\rm ^{12}C}_{\rm g.s.}). \label{Amp2}
\end{equation} 
As a result, we obtain the enhancement factor of 5 for the $\alpha$ spectroscopic 
factor of the 2$^+_1$ state of $^{12}$C compared to that of $^{12}$C$_\text{g.s.}$ 
\cite{Arima,Ku73,Volya17}. Given $S_\alpha\approx 0.794$ for the dissociation
$^{16}{\rm O}_{\rm g.s.}\to\alpha+^{12}$C$_{\rm g.s.}$ predicted by the large scale SM 
calculation \cite{Volya15,Volya17}, one finds from the relation (\ref{Amp2}) that 
$S_\alpha\approx 3.9$ for the dissociation $^{16}$O$_{\rm g.s.}\to\alpha+^{12}$C$_{2^+_1}$. 
This value was used as a fixed parameter in the present CRC calculation.  
Beside the SM results, the cluster \ac configurations of $^{16}$O have been studied 
also in the OCM by Suzuki \cite{Su76}, and recently in the 4$\alpha$ cluster OCM 
model by Yamada {\it et al.} \cite{Yama12}. The $S_\alpha$ values predicted 
by different structure models \cite{Volya15,Volya17,Yama12,Su76} for the 
$\alpha+^{12}$C$_{2^+_1}$ configuration are of 3 to 5 times that predicted 
for the $\alpha+^{12}$C$_\text{g.s.}$ configuration (see Table~\ref{tSex}), and this 
is clearly due to the degeneracy of the $M$-substates. For $^{12}$C in the $3^-_1$ 
state at 9.6 MeV, the total number of the oscillator quanta $\mathcal{N}$ for 
the configuration $\alpha+^{12}$C$_{3^-_1}$ is not conserved, because $N$ 
is implied by the Wildermuth's rule (\ref{crc3}) to be half-integer. Therefore,  
it cannot contribute to the $\alpha$ transfer cross section. Another important
state of the $^{12}$C core is the $0^+_2$ excitation at 7.65 MeV (Hoyle state), but 
$S_\alpha\approx 0.06$ predicted \cite{Ku73} for the dissociation 
$^{16}{\rm O}_{\rm g.s.}\to\alpha+^{12}$C$_{0^+_2}$ is too small to make any contribution
to the indirect $\alpha$ transfer. This result also agrees with the $\alpha$ pickup 
data \cite{Woz76,Oel78,Ume84} where $S_\alpha$ obtained for the $\alpha+^{12}$C$_{0^+_2}$ 
configuration is nearly 4 times smaller than that of the $\alpha+^{12}$C$_{\rm g.s.}$ 
configuration. In fact, the loosely bound 3$\alpha$ structure of the Hoyle state leads to 
a fragile $\alpha+^{12}$C$_{0^+_2}$ configuration in $^{16}$O that is strongly 
mixed with the $4^+_2$ state of $^{16}$O at the energy above 18 MeV \cite{Enyo17}, 
which is mainly of 4$\alpha$ structure and lying too high for the back coupling 
to the $\alpha$ transfer channel in the elastic \oc scattering. Therefore, we did 
not include the $0^+_2$ state of the $^{12}$C core in our CRC calculation.      
The complex inelastic $^{12}$C$_{\rm g.s.}\to ^{12}$C$_{2^+_1}$ form factor was 
also calculated in the generalized DFM \cite{Kho00}, using the $2^+_1$ transition 
density obtained in the Resonating Group Method (RGM) by Kamimura \cite{Kam81}. 
This choice of the $2^+_1$ transition density has been well tested in the earlier 
folding model analysis of the inelastic \ac scattering at different energies 
\cite{Kho08}.   
\begin{table}[bht]
	\caption{$\alpha$ spectroscopic factors predicted by different structure models 
for the dissociation $^{16}$O$\to\alpha+^{12}$C, where the $^{12}$C core is in  
its ground state and $2^+_1$ excited state at 4.44 MeV. The $S_\alpha$ values 
used in the present CRC analysis of the elastic \oc scattering (shown in the last 
row) were obtained in the large scale SM calculation \cite{Volya15,Volya17}.} 
	\label{tSex}\vspace{0.5cm}
	\begin{tabular}{|c|c|c|c|} \hline
		\multicolumn{4}{|c|}{ $S_\alpha$ }  \\ \hline
		Model & $\alpha+^{12}$C$_{\rm g.s.}$ & $\alpha+^{12}$C$_{2^+_1}$ & Reference \\ \hline  
		SM & 0.228 & 1.265 & \cite{Rot68} \\
		SM & 0.235 & 1.30 & \cite{Ku73}  \\
		SM  & 0.296 & 1.48 & \cite{Arima} \\
		OCM & 0.30 & 1.397 & \cite{Su76} \\
		$4\alpha$-OCM & 0.59 & 1.47 & \cite{Yama12} \\ 
		SM & 0.794 & 3.90 & \cite{Volya15,Volya17}  \\ \hline
	\end{tabular}
\end{table}

\begin{figure}\vspace*{-1cm}\hspace*{0cm}
	\includegraphics[width=0.9\textwidth]{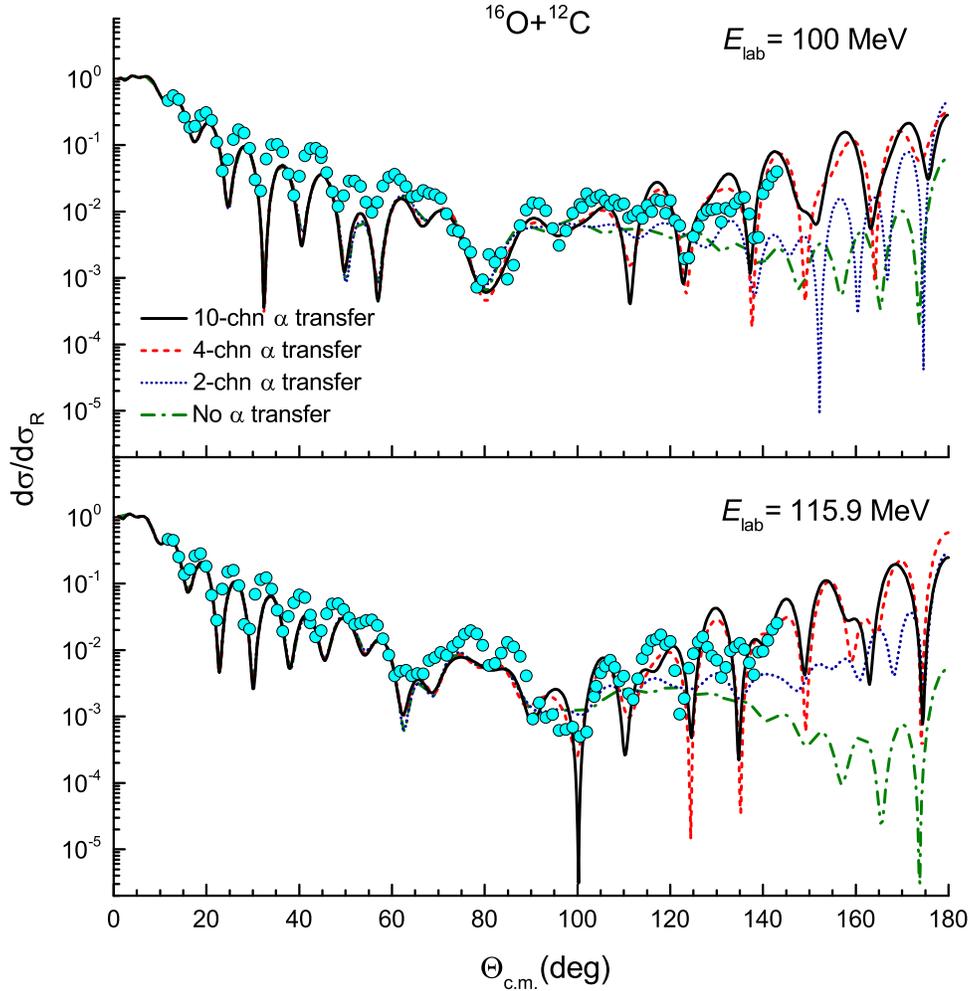}\vspace*{-1cm}
\caption{Full CRC description of the elastic \oc scattering data measured at 
$E_{\rm lab}=100$ and 115.9 MeV \cite{Nico00} (solid lines) in comparison with 
the CC results for purely elastic scattering, neglecting the $\alpha$ 
transfer (dash-dotted lines), the CRC results including the (direct) 2-channel $\alpha$ 
transfer (dotted lines), and the CRC results including the (direct and indirect via 
the $2^+_1$ state of the $^{12}$C core) 4-channel $\alpha$ transfer (dashed lines). 
See more details in text.} \label{fCRC1}
\end{figure}
\begin{figure}\vspace*{-1cm}\hspace*{0cm}
\includegraphics[width=0.9\textwidth]{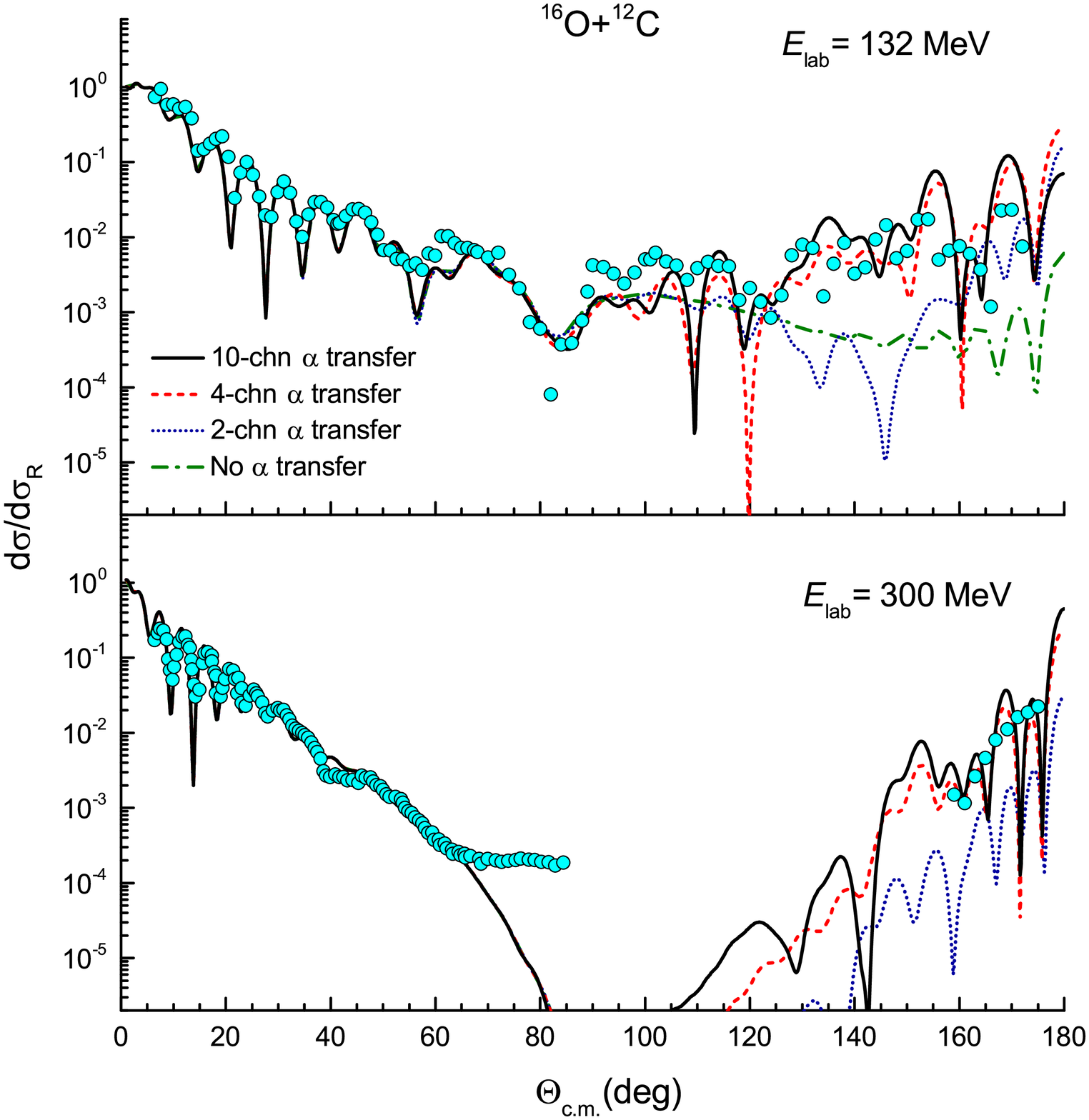}\vspace*{-1cm}
\caption{The same as Fig.~\ref{fCRC1} but for the elastic \oc scattering data measured
at $E_{\rm lab}=132$ MeV \cite{Oglo00,Oglo98}, and 300 MeV \cite{Bra01}.} \label{fCRC2}
\end{figure}
The total elastic \oc cross sections given by the present CRC calculation are compared 
with the data measured at $E_{\rm lab}=100-300$ MeV \cite{Nico00,Oglo00,Oglo98,Bra01} 
in Figs.~\ref{fCRC1} and \ref{fCRC2}. At variance with the CRC calculation of the direct 
$\alpha$ transfer discussed in Sect.~\ref{sec3b}, the 10-channel coupling shown in 
Fig.~\ref{fCoup} was found to affect the complex OP for the elastic \oc scattering 
significantly, especially, the coupling to the $2^+_1$ excitation of the $^{12}$C core
as found earlier in the folding model analysis of the inelastic \ac scattering 
\cite{Kho08}. The OP parameters were readjusted mainly for the best CRC fit to the 
elastic \oc data at forward angles which are of the true elastic scattering. One can 
see in Table~\ref{tOPind} that the absorption strength of the imaginary WS potential 
was reduced substantially at the considered energies due to the explicit coupling 
to the nonelastic channels shown in Fig.\ref{fCoup}. 
\begin{table}[bht]
\caption{The best-fit OP parameters (\ref{eOP}) used in the 10-channel CRC analysis 
of the elastic \oc scattering at $E_{\rm lab}=100-300$ MeV that includes both the direct
and indirect $\alpha$ transfers. All quantities are the same as those presented in 
Table~\ref{tOP}.} 	\label{tOPind} \vspace{0.5cm}
	\begin{tabular}{|c|c|c|c|c|c|c|c|} \hline
		$E_{\rm lab}$ & $N_R$ &    $J_R$     & $W_V$ & $R_V$ & $a_V$ & $J_W$ & $\sigma_R$ \\ 
		(MeV)   &   & (MeV~fm$^3$) & (MeV) & (fm)  & (fm) & (MeV~fm$^3$)  &  (mb) \\ \hline
		100     & 0.930 &  310.1   & 5.59  & 6.01  & 0.40 & 27.6  &    1331    \\ \hline
		115.9   & 0.910 &  301.6   & 5.50  & 6.00  & 0.40 & 27.1  &    1332    \\ \hline
		124     & 0.920 &  304.0   & 6.00  & 5.86  & 0.46 & 28.0  &    1347    \\ \hline
		132     & 0.945 &  311.3   & 7.60  & 5.99  & 0.55 & 38.5  &    1436    \\ \hline
		300     & 0.912 &  281.7   & 17.70 & 5.60  & 0.68 & 77.5  &    1551    \\ \hline  
	\end{tabular}
\end{table}
To reveal the contributions of different elastic $\alpha$ transfer paths to the 
elastic \oc cross section at large angles, we have performed several CRC calculations 
with different coupling schemes. The dash-dotted lines in Figs.~\ref{fCRC1} and \ref{fCRC2}
represent the results of the CC calculation of the elastic \oc scattering with only the
coupling to the excited states in the initial partition (a) included, neglecting
the $\alpha$ transfer. The dotted lines show the results of the CRC calculation 
including the direct elastic $\alpha$ transfer between the initial partition (a)
and its $\alpha$-exchanged counterpart (b). The dashed lines show the results 
of the CRC calculation including contributions from both the direct $\alpha$ 
transfer and indirect $\alpha$ transfer via the 2$^+_1$ state of the $^{12}$C core. 
Finally, the solid lines in Figs.~\ref{fCRC1} and \ref{fCRC2} show the results of the 
full CRC calculation including different $\alpha$-transfer paths between the considered 
10 channels. We note that the number of channels labeled in the figure legends 
corresponds to the number of reaction channels involving the $\alpha$ transfer 
between the two partitions (a) and (b). For each partition, the inelastic scattering 
channels were always coupled to the elastic scattering channel in all the CRC 
calculations discussed here. 

One can see in Figs.~\ref{fCRC1} and \ref{fCRC2} that without the contribution from 
the $\alpha$ transfer channels the CC calculation alone (dash-dotted lines) could 
describe the elastic \oc scattering at the forward- and medium angles only. A clear 
evidence of the failure of the CC method are the results shown in Fig.~\ref{fCRC2} 
for the elastic \oc scattering at $E_{\rm lab}=300$ MeV where the elastic scattering 
cross section falls exponentially with the increasing angles, and cannot be seen 
at all in the backward region. The present CC results thus show that the scenario 
suggested by Ohkubo and Hirabayashi \cite{Ohku14-2} for the large-angle oscillation 
of the elastic \oc cross section observed at $E_{\rm lab}=115.9$ MeV is not realistic. 
In fact, the CC calculation taking into account the couplings to the low-lying 
excitations of $^{16}$O and the $^{12}$C core cannot give rise to some ``reflective" 
wave that interferes with the elastic scattering wave at backward angles. As discussed 
above in Sect.~\ref{sec2}, it is not the CC effect but the use of a very small diffuseness 
of the WS imaginary potential in Ref.~\cite{Ohku14-2} that boosts the near-side 
scattering wave at backward angles for the interference with the far-side scattering 
wave, leading to the enhanced oscillation of the elastic cross section  
(as illustrated in Fig.~\ref{f116a}). 
 
With the direct and indirect $\alpha$ transfers taken into account, the elastic \oc
cross section is indeed enhanced at large angles. From the CRC results shown in 
Figs.~\ref{fCRC1} and \ref{fCRC2} one can see that the indirect $\alpha$ transfer is 
vital for a proper description of the enhanced oscillation of the elastic \oc cross 
section at backward angles. Our CRC results also show that the indirect $\alpha$ 
transfer through the 2$^+_1$ excitation of the $^{12}$C core is the dominant transfer 
channel (see, in particular, the CRC results for the elastic \oc scattering 
at $E_{\rm lab}=300$ MeV). This is naturally explained by the large $S_\alpha$ 
value predicted for the $\alpha+^{12}$C$_{2^+_1}$ configuration in the dissociation 
of $^{16}$O (see Table~\ref{tSex}). The coupling effects to the $\alpha$ transfer 
by the $0^+_2$, $3^-_1$, and $2^+_1$ excitations of $^{16}$O are relatively weak 
and can only be slightly seen in the 300 MeV cross section. The 300 MeV data points 
at the most backward angles should be a good reference for gauging the strength 
of the $\alpha$ spectroscopic factor and probing the impact by the $\alpha$ transfer 
on the elastic \oc scattering.
Like the results obtained earlier by Rudchik {\it et al.} \cite{Rud10}, the use 
of the small $\alpha$ spectroscopic factors given by the traditional (old) definition 
of $S_\alpha$ in our CRC calculation gives a much smaller elastic $\alpha$ 
transfer cross section cross section at backward angles, leaving the large-angle 
elastic \oc data unexplained at the considered energies. Therefore, the present 
CRC results for the elastic \oc scattering seems to support the use of the \emph{new} 
definition of the $\alpha$ spectroscopic factor \cite{Volya15,Kra17}. 
In this connection, more theoretical and experimental studies of the $\alpha$ transfer 
and knock-out reactions are highly desired for the systematic and reliable information 
on the spectroscopic properties of the $\alpha$-cluster dissociation of light nuclei. 

\begin{figure}\vspace*{-1cm}\hspace*{0cm}
\includegraphics[width=0.9\textwidth]{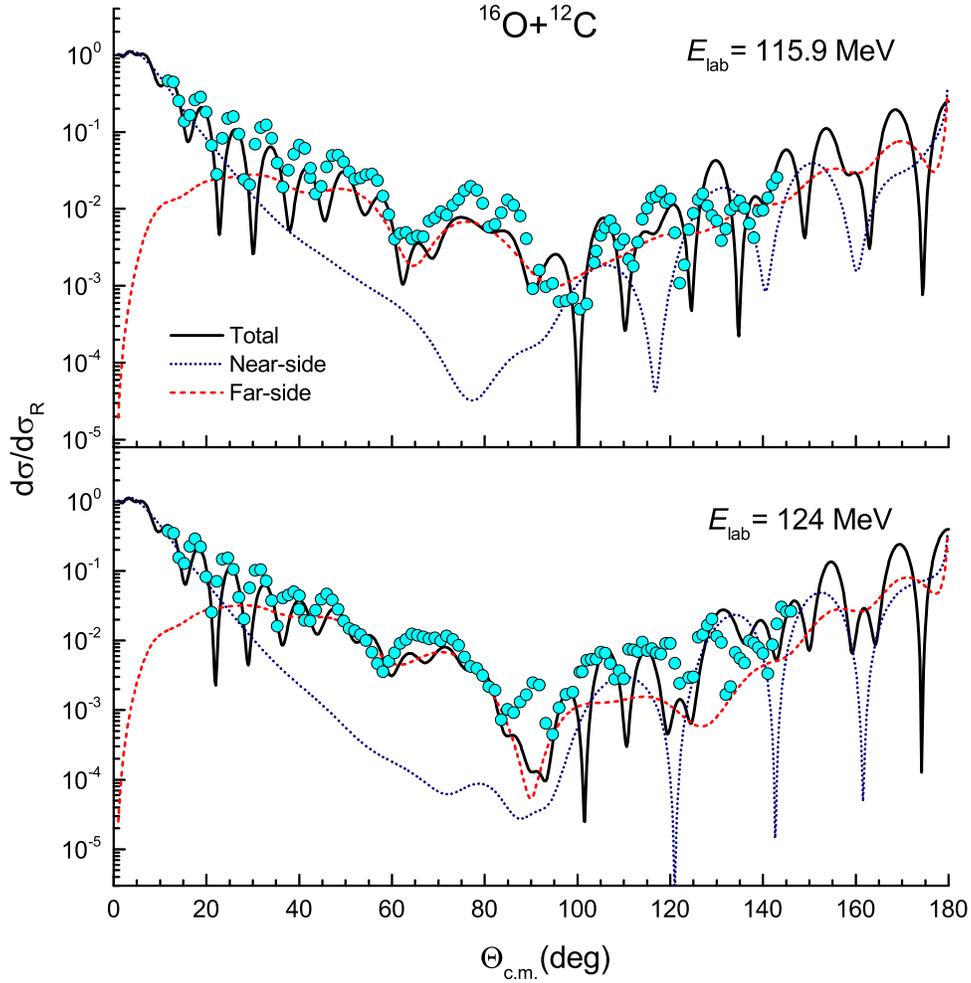}\vspace*{-1cm}
 \caption{10-channel CRC description (solid lines) of the elastic \oc data measured 
at $E_{\rm lab}=115.9$ and 124 MeV \cite{Nico00}. The total elastic \oc cross 
section is decomposed into the near-side (dotted lines) and far-side (dashed lines) 
contributions (\ref{NFdec}) using Fuller's method \cite{Ful75}.} \label{fNFx}
\end{figure}
As discussed above in Sect.~\ref{sec2}, the enhanced oscillation of the elastic 
\oc cross section observed at backward angles could be reproduced in the conventional 
OM calculation only if a very small diffuseness of the WS imaginary OP is used.
Such an unusual WS imaginary OP enhances the near-side scattering at large angles, 
and gives rise, therefore, to a strong near-far interference there. In general, 
the near-side component of the elastic \AA scattering is dominant in the surface 
region, at forward angles only. The large-angle scattering (if not suppressed 
by the strong absorption) is refractive and mainly of the far-side strength 
\cite{Kho07r,Bra96,Frahn84}. Therefore, a scenario for the strong near-side 
surface scattering at backward angles is very unlikely.     
Given the strong impact of the $\alpha$ transfer on the elastic \oc scattering 
at large angles shown in Figs.~\ref{fCRC1} and \ref{fCRC2}, it is of interest 
to explore whether the $\alpha$ transfer process also enhances the near-side 
scattering at backward angles. For this purpose, the total elastic amplitude 
$f(\theta)$ given by the full (10-channel) CRC calculation has been decomposed 
into the near-side and far-side components (\ref{NFdec}) using the Fuller's method 
\cite{Ful75}, and the results are shown in Fig.~\ref{fNFx}. One can see that 
the elastic $\alpha$ transfer indeed enhances the strength of the near-side 
scattering at backward angles and, in the same way as shown in the OM results 
in Fig.~\ref{f116a}, the near-far interference does give rise to the enhanced 
oscillation of the elastic \oc cross section at backward angles. However, the 
enhanced near-side cross section at large angles shown in Fig.~\ref{fNFx} is 
\emph{not} of the true elastic scattering but caused by the elastic $\alpha$ 
transfer that occurs mainly at the surface, at forward angles. The simple reason why 
it shows up in the elastic cross section at backward angles is that the $\alpha$ 
transfer amplitudes at $(\pi-\theta)$ were added to the elastic scattering 
amplitude at $\theta$, using the relation (\ref{fsig}). Therefore, the enhanced 
near-side cross section at large angles shown in Fig.~\ref{fNFx} is not an 
artificial effect caused by a specific numerical technique of the OM calculation, 
but originates naturally from the $\alpha$ transfer process. The CRC results 
were obtained using the \oc real folded potential and WS imaginary potential 
with a normal diffuseness $a_V\approx 0.4-0.6$ fm (see Table~\ref{tOPind}). 
We conclude, therefore, that the unusual imaginary WS potential having 
a very small diffuseness deduced from the original OM analysis of these 
data \cite{Nico00} probably mimics the dynamic polarization of the OP 
by a strong coupling between the true elastic \oc scattering channel and different 
$\alpha$ transfer channels.  

\section{Impact by other transfer channels}
It was shown in Ref.~\cite{Rud10} that the two-step transfer processes such 
as the nucleon transfer or sequential transfers of a neutron (proton) and 
$^{3}$He (triton) also contribute to the enhanced backward-angle oscillation 
of the elastic \oc cross section at low energies. Therefore, it is also 
of interest for the present research to investigate explicitly the contribution 
from such transfer processes, to see if they can alter the important 
conclusion made above on the impact by $\alpha$ transfer. 

In the CRC analysis of the mentioned transfer channels, the real optical potentials 
for all the outgoing channels were calculated in the DFM using the g.s. densities 
of $^{11}$B, $^{11,13}$C, $^{13,15}$N, $^{15,17}$O, and $^{17}$F given by the 
independent particle model (IPM) \cite{Sat79}. The same WS imaginary OP was assumed
for the entrance and exit channels, with the parameters given in Table~\ref{tOP} 
for each energy. The nucleon- and ($A=3$) cluster binding potentials were chosen 
in the WS form with the fixed radius $r_0=1.25$ fm and diffuseness $a=0.65$ fm, 
while the WS depths were adjusted to reproduce the observed nucleon- and ($A=3$) 
cluster separation energies. The nucleon spectroscopic factors were taken 
from the compilation by Tsang {\it et al.} \cite{Tsang} which is based on a 
systematic DWBA analysis of $(d,p)$ reactions, and also in a good agreement with 
the SM results. For simplicity, the isospin symmetry was adopted for the neutron 
and proton spectroscopic factors. Due to the lack of the experimental data for $^{3}$He 
and triton spectroscopic factors, we have used the values predicted by the SM as 
quoted in Ref.~\cite{Rud10}.

\subsection{Nucleon transfer reaction}
\begin{figure}\vspace*{0cm}\hspace*{0cm}
\includegraphics[width=0.9\textwidth]{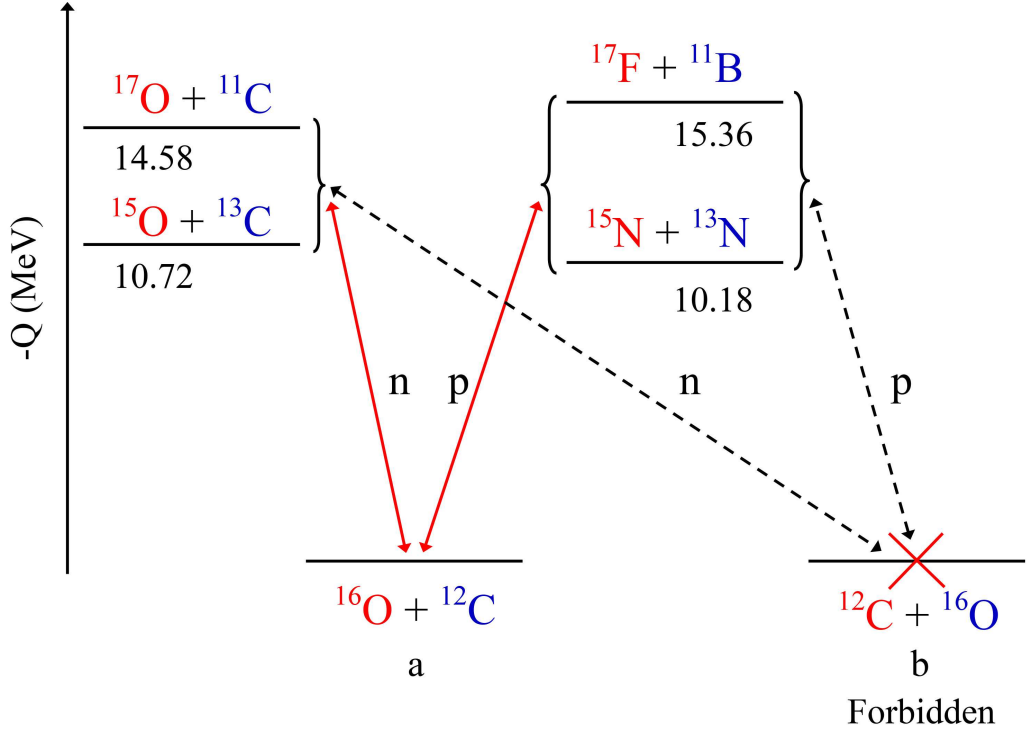}\vspace*{0cm}
\caption{Explicit coupling between the true elastic scattering and nucleon
transfer channels taken into account in the CRC calculation of the elastic 
\oc scattering. The $\alpha$-exchanged partition (b) of the \oc system cannot 
be populated by the nucleon transfer reaction.} \label{fNtran}
\end{figure}
The strength of the coupled channel contribution from the nucleon transfer 
reactions to the elastic \oc cross section can be determined from the CRC calculation 
based on the coupling scheme shown in Fig.~\ref{fNtran}. It is important to note 
that the two-step (back and forth) nucleon transfer proceeds through the two-way 
couplings between the initial partition (a) and 4 corresponding nucleon-exchanged 
partitions shown in Fig.~\ref{fNtran}, and the CRC equations similar to 
Eqs.~(\ref{crc1})-(\ref{crc2}) are solved by an iterative procedure \cite{Tho09,Tho88}. 
The total elastic scattering amplitude is then calculated from the \emph{converged} 
CRC wave function of the elastic scattering channel. It is obvious from Fig.~\ref{fNtran} 
that the $\alpha$-exchanged partition (b) of the \oc system cannot be populated by the 
nucleon transfer reactions.  
\begin{figure}\vspace*{-1.5cm}\hspace*{0cm}
\includegraphics[width=0.9\textwidth]{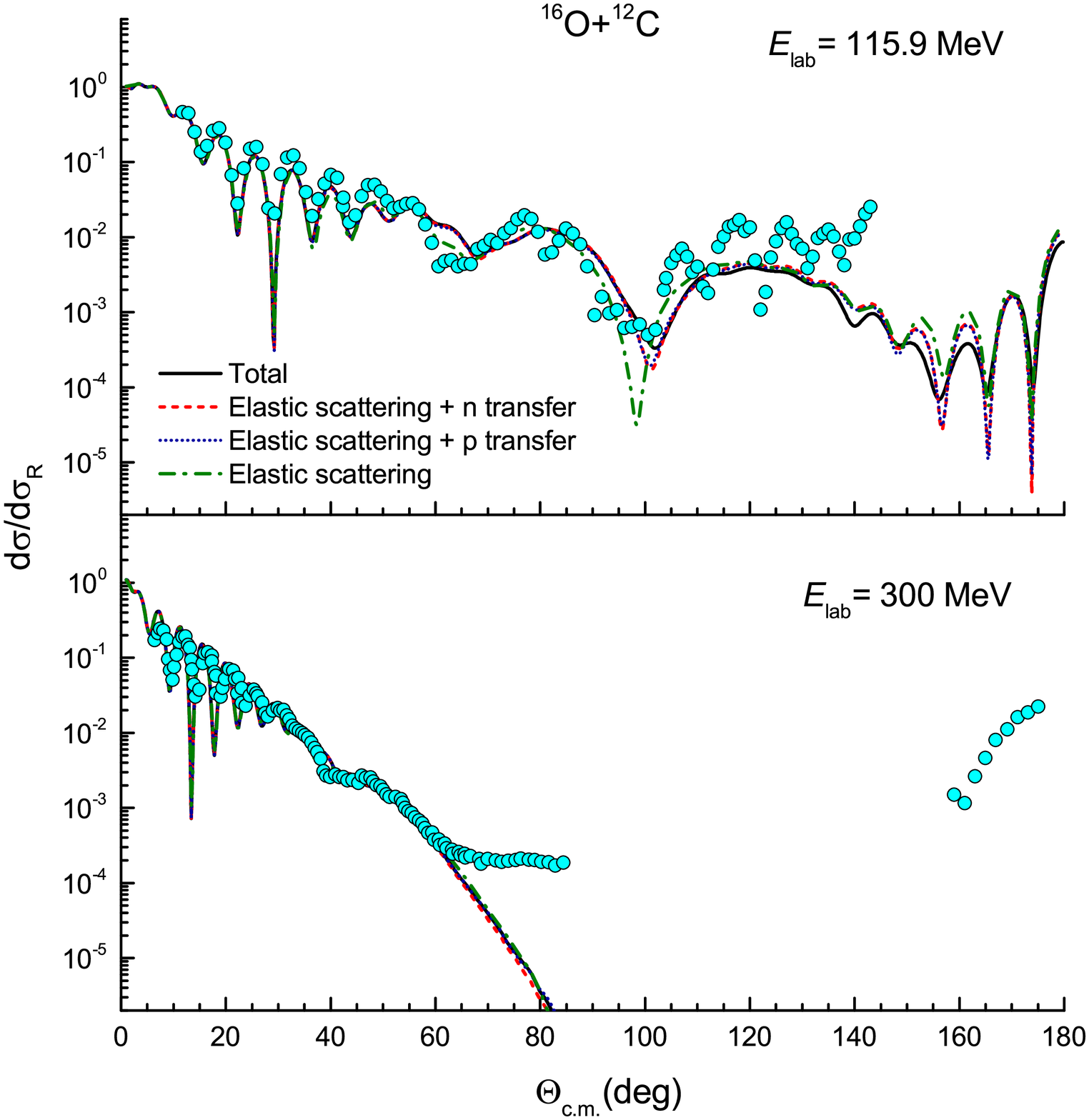}\vspace*{-1cm}
\caption{CRC description of the elastic \oc scattering data measured at 
$E_{\rm lab}=115.9$ \cite{Nico00} and 300 MeV \cite{Bra01} based on the coupling
scheme shown in Fig.~\ref{fNtran}. The solid lines are the results given by the 
full coupling between the true elastic scattering and all nucleon transfer channels. 
The CRC results obtained separately with the coupling to the neutron- or proton 
transfer channel are shown as the dashed and dotted lines, respectively. The dash-dotted 
lines are the results of the OM calculation alone.} 
 \label{fXNtr}
\end{figure}

One can see from the CRC results shown in Fig.~\ref{fXNtr} that the neutron and 
proton transfer channels have a minor coupled channel effect on the elastic \oc
scattering cross section at the low energy of 115.9 MeV, and is completely 
negligible at the higher energy of 300 MeV. Moreover, the nucleon transfer channels 
do not contribute at all to the formation of the oscillatory pattern of the elastic 
cross section at large angles. Because the $\alpha$-exchanged partition (b) of the 
\oc system cannot be populated by the nucleon transfer as shown in Fig.\ref{fNtran},
the two-step (back and forth) nucleon transfer contributes to the elastic scattering
cross section only through the small change of this cross section induced by the 
coupled channels effect (shown in the upper panel of Fig.\ref{fXNtr}). 
The CRC calculation does not generate separately the elastic two-step nucleon 
transfer cross sections, denoted in Ref.~\cite{Rud10} as $\langle n,n\rangle$ 
and $\langle p,p\rangle$, that can be added to the elastic scattering cross section. 
In fact, only if one includes the \emph{forbidden} two-step nucleon transfer 
to the $\alpha$-exchanged partition (b) of the \oc system, then there appear 
4 more CRC equations that lead separately to the two-step $\langle n,n\rangle$ and 
$\langle p,p\rangle$ transfer cross sections as discussed in Ref.~\cite{Rud10}. 
As a test, we did such a CRC calculation and obtained about the same $\langle n,n\rangle$ 
and $\langle p,p\rangle$ cross sections as those shown in the right panel of Fig.~7 
of Ref.~\cite{Rud10}, which are still nearly 2 orders of magnitude smaller than the 
elastic scattering cross section at medium and backward angles. We conclude, therefore, 
that the nucleon transfer process cannot be the source of the enhanced oscillation 
of the elastic \oc cross section at large angles.  

\subsection{Sequential transfers of the neutron-$^{3}$He and proton-triton pairs}
\begin{figure}\vspace*{-0cm}\hspace*{0cm}
	\includegraphics[width=0.8\textwidth]{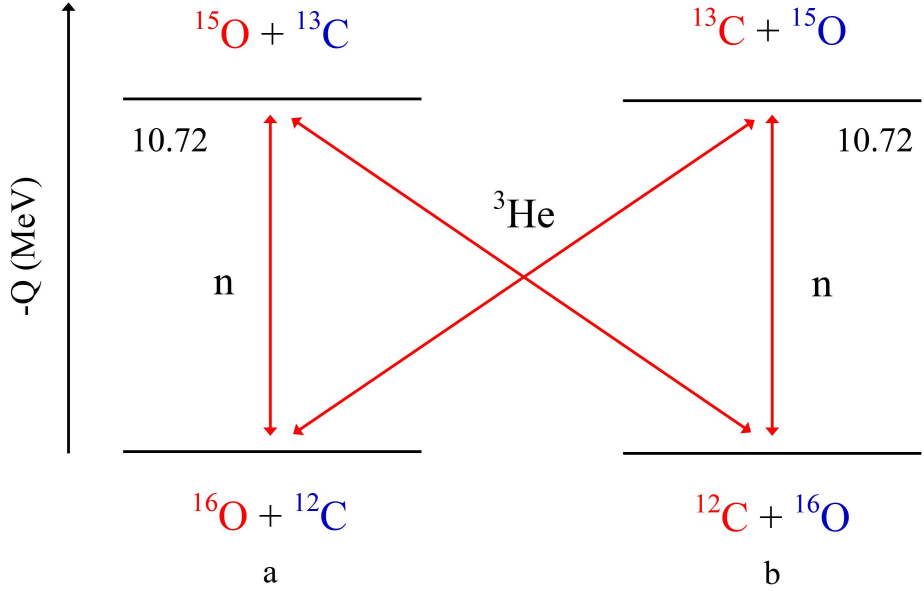}\vspace*{-0.5cm}
 \caption{Two-way coupling between the true elastic scattering and sequential
 $\langle n,^{3}$He$\rangle$  transfer channels taken into account in the CRC 
calculation of the elastic \oc scattering. The coupling scheme for the sequential 
$\langle p,^{3}$H$\rangle$  transfer via the $^{15}$N+$^{13}$N partitions 
at $-Q=10.18$ MeV is exactly the same. } \label{f3tran}
\end{figure}

Another possible source for the backward-angle enhancement of the elastic \oc 
cross section is the sequential two-step transfer of the neutron-$^{3}$He or 
proton-triton pair \cite{Rud10} that naturally populates the $\alpha$-exchanged 
partition (b) of the \oc system via the two-way coupling scheme shown, e.g., 
for the $\langle n,^{3}$He$\rangle$ transfer in Fig.~\ref{f3tran}. 

\begin{figure}\vspace*{-1cm}\hspace*{0cm}
\includegraphics[width=0.9\textwidth]{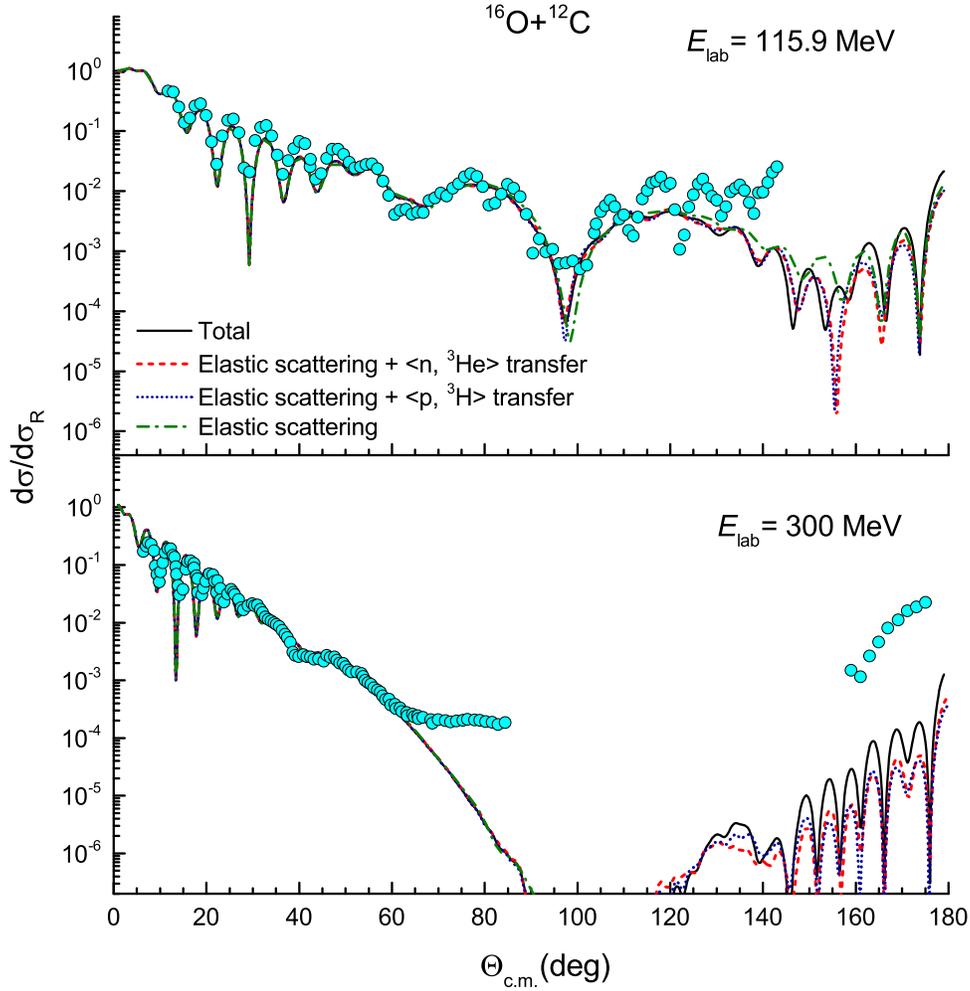}\vspace*{-1cm}
\caption{CRC description of the elastic \oc scattering data measured at 
$E_{\rm lab}=115.9$ \cite{Nico00} and 300 MeV \cite{Bra01} based on the coupling
scheme shown in Fig.~\ref{f3tran}. The solid lines are the results given by the 
full coupling between the true elastic scattering and sequential two-step 
$\langle n,^{3}$He$\rangle$ and $\langle p,^{3}$H$\rangle$ transfer channels. 
The CRC results obtained separately with the coupling to the $\langle n,^{3}$He$\rangle$ 
or $\langle p,^{3}$H$\rangle$ transfer channel are shown as the dashed 
and dotted lines, respectively. The dash-dotted lines are the results of the OM 
calculation alone.} \label{f3Ntr}
\end{figure}

The results of the CRC calculation based on the coupling scheme illustrated in 
Fig.~\ref{f3tran} are shown in Fig.~\ref{f3Ntr}. One can see that the sequential 
two-step transfers of the $n-^{3}$He and $p-^{3}$H pairs do have some impact on 
the elastic \oc cross section at backward angles. At the low energy of 115.9 MeV, 
these transfer processes have only a minor effect on the backward-angle oscillation,
while at 300 MeV they give rise to an enhanced oscillation of the elastic cross 
section at backward angles, but the calculated cross section still underestimates 
the data by about two orders of magnitude. 
These results of our CRC analysis of the sequential two-step $\langle n,^{3}$He$\rangle$ 
and $\langle p,^{3}$H$\rangle$ transfers to the $\alpha$-exchanged partition (b) show 
clearly that they cannot contribute significantly to the observed oscillation of the 
elastic \oc cross section at backward angles. Such effect is clearly due to the direct 
and indirect $\alpha$ transfer processes discussed in Sect.~\ref{sec3c}. 

\section*{Summary}
The enhanced backward-angle oscillation of the elastic \oc scattering cross section 
observed at low energies was shown to distort strongly the smooth pattern of the 
refractive (rainbow) scattering established for this system. To explore this effect, 
we have performed a detailed CRC analysis of the elastic \oc scattering, taking 
explicitly into account the coupling between the elastic scattering and different 
$\alpha$ transfer channels, using the folded \oc and \cc potentials as the real 
optical potentials for these systems. The direct (elastic) $\alpha$ transfer alone 
was found to account properly for the enhanced oscillation of the elastic cross 
section at backward angles only by using a best-fit $\alpha$ spectroscopic factor 
that is much larger than the $S_\alpha$ values predicted by different structure 
models. 

The disagreement of the best-fit $S_\alpha$ deduced from the CRC analysis of the 
direct $\alpha$ transfer and those predicted by the structure calculations lead us
to consider the indirect $\alpha$ transfer channels. With the elastic scattering 
channel coherently coupled to the inelastic scattering channels as well as to the 
direct and indirect (via the low-lying excitations of $^{16}$O and the $^{12}$C core) 
$\alpha$ transfer channels in the present CRC calculation, a satisfactory description 
of the considered elastic \oc data has been achieved, using the $\alpha$ spectroscopic 
factors predicted by the large scale SM calculation (the cluster-nucleon configuration 
interaction model by Volya and Tchuvilsky \cite{Volya15,Volya17}). The indirect 
$\alpha$ transfer via the 2$^+_1$ excitation of the $^{12}$C core was found to be 
the dominant $\alpha$ transfer channel. Thus, our CRC results seem to support, for the 
first time in the direct reaction studies, the use of the \emph{new} microscopic 
definition of the $\alpha$ spectroscopic factor \cite{Volya15,Kra17}.  

The decomposition of the total elastic amplitude into the near-side and far-side 
components using the Fuller's method \cite{Ful75} allowed us to conclude that the
enhanced backward-angle oscillation of the elastic \oc cross section observed
at low energies is caused by the interference between the near-side and far-side
scattering waves. While the far-side scattering wave is of the true elastic 
\oc scattering, the unusually strong near-side scattering wave at backward angles 
is caused by the $\alpha$ transfer, and shows up in the elastic cross section 
because of the identity of the initial and $\alpha$-exchanged partitions.    

For the completeness of the present study, the coupled channel effects to the
elastic \oc scattering at low energies from other two-step transfer channels, 
like the nucleon transfer and sequential transfers of a neutron-$^{3}$He and 
proton-triton pairs, were investigated in detail and their contribution 
to the elastic \oc cross section at backward angles were found negligible. 
The enhanced backward-angle oscillation of the elastic \oc cross section 
at low energies is, therefore, mainly due to direct and indirect $\alpha$ 
transfer processes    

The strong coupling effect by the $\alpha$ transfer channels found in the 
present work is the motivation for a consistent CRC study of the $\alpha$ transfer 
in both the elastic and inelastic \oc scattering at low energies, which is 
planned as the follow-up research.   

\section*{Acknowledgments}
The present research has been supported by the National Foundation for 
Scientific and Technological Development (NAFOSTED Project No. 103.04-2016.35). 
The authors also thank N. Keeley and A. Moro for their helpful communication
on the inputs of the CRC calculation with the code FRESCO, A. Volya for his 
comments on the microscopic determination of the $\alpha$ spectroscopic 
factor. S. Okabe and M. Kamimura are much appreciated for providing us with the 
nuclear densities obtained in the OCM and RGM calculations, respectively. 
The IPM densities were calculated using the code DOLFIN provided to one 
of us (D.T.K.) by the late Ray Satchler.


\begin{thebibliography}{99}
\bibitem{Sa79} G.R. Satchler and W.G. Love, Phys. Rep. {\bf 55}, 183 (1979).
\bibitem{Bra97} M.E. Brandan and G.R. Satchler, Phys. Rep. {\bf 285}, 143 (1997).
\bibitem{Kho07r} D.T. Khoa, W. von Oertzen, H.G. Bohlen, and S. Ohkubo,
 J. Phys. G {\bf 34}, R111 (2007).
\bibitem{Kho16} D.T. Khoa, N.H. Phuc, D.T. Loan, and B.M. Loc,
	Phys. Rev. C {\bf 94}, 034612 (2016).
\bibitem{Bra91} M.E. Brandan and G.R. Satchler, Phys. Lett. B {\bf 256}, 311 (1991).
\bibitem{Rou85} P. Roussel, N. Alamanos, F. Auger, J. Barrette, B. Berthier, 
 B. Fernandez, L. Papineau, H. Doubre, and W. Mittig, Phys. Rev. Lett. {\bf 54}, 1779 (1985).
\bibitem{Bra86} M.E. Brandan, A. Menchaca-Rocha, M. Buenerd, J. Chauvin, 
 P. De Saintignon, G. Duhamel, D. Lebrum, P. Martin, G. Perrin, and J.Y. Hostachy, 
 Phys. Rev. C {\bf 34}, 1484 (1986). 
\bibitem{Vil89} A.C.C. Villari, A. L\'{e}pine-Szily, R.L. Filho, O.P. Filho, M.M. Obuti, 
 J.M. Oliveira Jr, and N. Added, Nucl. Phys. A  {\bf 501}, 605 (1989).
\bibitem{Oglo98} A.A. Ogloblin, D.T. Khoa, Y. Kond\=o, Yu.A. Glukhov, 
 A.S. Dem’yanova, M.V. Rozhkov, G.R. Satchler, and S.A. Goncharov, 
 Phys. Rev. C {\bf 57}, 1797 (1998). 
\bibitem{Oglo00} A.A. Ogloblin, Yu.A. Glukhov, W.H. Trzaska, A.S. Demyanova, 
	S.A. Goncharov, R. Julin, S.V. Klebnikov, M. Mutterer, M.V. Rozhkov, 
	V.P. Rudakov, G.P. Tiorin, D.T. Khoa, and G.R. Satchler, 
	Phys. Rev. C {\bf 62}, 044601 (2000).
\bibitem{Nico00} M.P. Nicoli, F. Haas, R.M. Freeman, S. Szilner, Z. Basrak, A. Morsad, 
 G.R. Satchler, and M. E. Brandan, Phys. Rev. C  {\bf 61}, 034609 (2000).
\bibitem{Bra01} M.E. Brandan, A. Menchaca-Rocha, L. Trache, H.L. Clark,
	A. Azhari, C. A. Gagliardi, Y.-W. Lui, R. E. Tribble, R. L. Varner,
	J. R. Beene, and G. R. Satchler, Nucl. Phys. A {\bf 688}, 659 (2001).
\bibitem{Kho94} D.T. Khoa, W. von Oertzen, and H.G. Bohlen, 
 Phys. Rev. C {\bf 49}, 1652 (1994).
\bibitem{Kho97} D.T. Khoa, G.R. Satchler, and W. von Oertzen,
 Phys. Rev. C {\bf 56}, 954 (1997).
\bibitem{Bra88} M.E. Brandan and G.R. Satchler, Nucl. Phys. A {\bf 487}, 477 (1988).
\bibitem{Kho00a} D.T. Khoa, W. von Oertzen, H.G. Bohlen, and F. Nuoffer, 
	Nucl. Phys. A {\bf 672}, 387 (2000).
\bibitem{Mich01} F. Michel, G. Reidemeister, and S. Ohkubo, Phys. Rev. C {\bf 63}, 034620 (2001).
\bibitem{Ohku14-1} S. Ohkubo and Y. Hirabayashi, Phys. Rev. C {\bf 89}, 051601(R) (2014).
\bibitem{Brau82} P. Braun-Munziger and J. Barette, Phys. Rep. {\bf 87}, 209 (1982).
\bibitem{vOe75} W. von Oertzen and H.G. Bohlen, Phys. Rep. {\bf 19 C}, 1 (1975).
\bibitem{Szi02} S. Szilner, W. von Oertzen, Z. Basrak, F. Haas, and M. Milin,
	Eur. Phys. J. A {\bf 13}, 273 (2002). 
\bibitem{Mor11} M.C. Morais and R. Lichtenth\"{a}ler, Nucl. Phys. A {\bf 857}, 1 (2011).
\bibitem{Hama11} Sh. Hamada, N. Burtebayev, K.A. Gridnev, and N. Amangeldi, 
 Nucl. Phys. A {\bf 859}, 29 (2011). 
\bibitem{Rud10} A.T. Rudchik {\it et al.}, Eur. Phys. J. A {\bf 44}, 221 (2010).
\bibitem{Ohku14-2} S. Ohkubo and Y. Hirabayashi, Phys. Rev. C {\bf 89}, 061601(R) (2014).
\bibitem{Arima} M. Ichimura, A. Arima, E.C. Halbert, and T. Terasawa, 
 Nucl. Phys. A {\bf 204}, 225 (1973).
\bibitem{Sat83} G.R. Satchler, {\it Direct Nuclear Reactions} (Clarendon, Oxford, 1983).
\bibitem{Tho09} I.J. Thompson and F.M. Nunes, {\it Nuclear Reactions for Astrophysics} 
(Cambridge University Press, Cambridge, UK, 2009).
\bibitem{Volya15} A. Volya and Y.M. Tchuvilsky, Phys. Rev. C {\bf 91}, 044319 (2015).
\bibitem{Yama12} T. Yamada, Y. Funaki, T. Myo, H. Horiuchi, K. Ikeda, G. R\"opke,
 P. Schuck, and A. Tohsaki, Phys. Rev. C {\bf 85}, 034315 (2012).
\bibitem{Volya17} A. Volya, private communication (unpublished). 
\bibitem{Kho00} D.T. Khoa and G.R. Satchler, Nucl. Phys. A {\bf 668}, 3 (2000). 
\bibitem{Pol76} J.E. Poling, E. Norbeck, and R.R. Carlson, Phys. Rev. C {\bf 13}, 648 (1976).
\bibitem{Kho01} D.T. Khoa, Phys. Rev. C {\bf 63}, 034007 (2001).
\bibitem{Raynal} J. Raynal, {\it Computing as a Language of Physics}
 (IAEA, Vienna, 1972) p.~75;  J. Raynal, coupled-channel code ECIS97 (unpublished).
\bibitem{Ful75} R.C. Fuller, Phys. Rev. C {\bf 12}, 1561 (1975).
\bibitem{Bra96} M.E. Brandan, M.S. Hussein, K.W. McVoy, and G.R. Satchler, 
	{\it Comments on nuclear and particle physics}, Vol.~22 (Gordon and Breach, 
	New York, 1996), p.~77. 
\bibitem{Frahn84} W.E. Frahn, {\it Treaties on Heavy-Ion Science} vol.~1, p.~135, 
 ed. D.A. Bromley (Plenum Press, New York, 1984).
\bibitem{Tho88} I.J. Thompson, Comput. Phys. Rep. {\bf 7}, 167 (1988); 
 http://www.fresco.org.uk.
\bibitem{Kho08} D.T. Khoa and D.C. Cuong, Phys. Lett. B {\bf 660}, 331 (2008). 
\bibitem{Til93} D.R. Tilley, H.R. Weller, and C.M. Cheves, Nucl. Phys. A {\bf 564}, 1 (1993).
\bibitem{Kra17} K. Kravvaris and A. Volya, Phys. Rev. Lett. {\bf 119}, 062501 (2017).
\bibitem{Flis76} T. Fliessbach and H.J. Mang, Nucl. Phys. A {\bf 263}, 75 (1976).
\bibitem{Flis77} T. Fliessbach and P. Manakos, J. Phys. G {\bf 3}, 643 (1977).
\bibitem{Lovas} R.G. Lovas, R.J. Liotta, A. Insolia, K. Varga, and D.S. Delion, 
 Phys. Rep. {\bf 294}, 265 (1998).
\bibitem{Betan} R. Id Betan and W. Nazarewicz, Phys. Rev. C {\bf 86}, 034338 (2012).
\bibitem{Lov85} R.G. Lovas, Z. Phys. A {\bf 322}, 589 (1985).
\bibitem{Grid13} K.A. Gridnev, N.A. Maltsev, and N.V. Leshakova, 
 Bull. Russ. Acad. Sci. Phys. {\bf  77}, 852 (2013).
\bibitem{Hama14} Sh. Hamada, N. Burtebayev, and N. Amangeldi, 
 Int. J. Mod. Phys. E {\bf 23}, 1450061 (2014).
\bibitem{Enyo17} Y. Kanada-En'yo, Phys. Rev. C {\bf 96}, 034306 (2017).
\bibitem{Okabe} S. Okabe, {\it Tours Symposium on Nuclear Physics II}, edited
 by H. Utsunomiya {\it et al.} (World Scientific, Singapore, 1995) p.~112, 
 and private communications.
\bibitem{Miska} H. Miska, H.D. Gr\"{a}f, A. Richter, R. Schneider, D. Sch\"{u}ll, 
E. Spamer, H. Theissen, O. Titze, and Th.Walcher, Phys. Lett. B {\bf 58}, 155 (1975).
\bibitem{Su76} Y. Suzuki, Prog. Theor. Phys. {\bf 55}, 1751 (1976); {\bf 56}, 111 (1976).
\bibitem{Rot68} I. Rotter, Fortschritte der Physik {\bf 16}, 195 (1968).
\bibitem{Ku73} D. Kurath, Phys. Rev. C {\bf 7}, 1390 (1973).	
\bibitem{Woz76} G.J. Wozniak, D.P. Stahel, J. Cerny, and N.A. Jelley, 
 Phys. Rev. C {\bf 14}, 815 (1976).
\bibitem{Oel78} W. Oelert, A. Djaloeis, C.Mayer-B\"{o}ricke, P. Turek, and S. Wiktor, 
 Nucl. Phys. A {\bf 306}, 1 (1978).
\bibitem{Ume84} K. Umeda {\it et al.}, Nucl. Phys. A {\bf 429}, 88 (1984).
\bibitem{Kam81} M. Kamimura, Nucl. Phys. A {\bf 351}, 456 (1981).
\bibitem{Sat79} G.R. Satchler, Nucl. Phys. A {\bf 329}, 233 (1979).
\bibitem{Tsang} M.B. Tsang, J. Lee, and W.G. Lynch, 
 Phys. Rev. Lett. {\bf 95}, 222501 (2005).
\end{thebibliography}
\end{document}